\documentclass[fleqn,usenatbib]{mnras}

\usepackage{newtxtext,newtxmath}
\usepackage[T1]{fontenc}

\usepackage{tikz,xcolor,hyperref}
\definecolor{lime}{HTML}{A6CE39}
\DeclareRobustCommand{\orcidicon}{%
	\begin{tikzpicture}
	\draw[lime, fill=lime] (0,0) 
	circle [radius=0.16] 
	node[white] {{\fontfamily{qag}\selectfont \tiny ID}};
	\draw[white, fill=white] (-0.0625,0.095) 
	circle [radius=0.007];
	\end{tikzpicture}
	\hspace{-2mm}
}
\foreach \x in {A, ..., Z}{%
	\expandafter\xdef\csname orcid\x\endcsname{\noexpand\href{https://orcid.org/\csname orcidauthor\x\endcsname}{\noexpand\orcidicon}}
}
\newcommand{\teff}{\mbox{$T_{\rm eff}$}}

\newcommand{\numax}{\mbox{$\nu_{\rm max}$}}

\newcommand{\logg}{\mbox{$\log g$}}
\newcommand{\kep}{\mbox{\textit{Kepler}}}

\newcommand{\msun}{\mbox{$\rm M_{\odot}$}}
\newcommand{\logl}{\mbox{$\rm \log (L/L_{\odot})$}}

\newcommand{\kwthree}{\mbox{$K_{\rm s}$\textendash [12]}}
\newcommand{\kwfour}{\mbox{$K_{\rm s}$\textendash [22]}}
\newcommand{\mk}{\mbox{$M_{\rm K_s}$}}
\newcommand{\redbf}[1]{\textcolor{black}{\textnormal{{#1}}}}

\defcitealias{yu2020a}{Paper I}
\usepackage{graphicx}	% Including figure files
\usepackage{amsmath}	% Advanced maths commands
\usepackage{natbib}

\title[Mass loss in luminous red giants]{Asteroseismology of luminous red giants with {\it Kepler}. II. \redbf{Dependence of mass loss on pulsations and radiation}}

\author[Yu et al.]{
Jie Yu\orcidJ{}$^{1}$\thanks{Corresponding author: yujie@mps.mpg.de (JY)},
Saskia Hekker\orcidS{}$^{2,3}$,
Timothy R. Bedding\orcidT{}$^{4,5}$,
Dennis Stello\orcidD{}$^{6,4,5}$,
\newauthor
Daniel Huber\orcidK{}$^{7}$,
Laurent Gizon\orcidG{}$^{1,8,9}$,
Shourya Khanna\orcidK{}$^{10}$,
and Shaolan Bi\orcidB{}$^{11}$
\\
$^{1}$Max Planck Institute for Solar System Research, Justus-von-Liebig-Weg 3, 37077 Göttingen, Germany\\
$^{2}$Heidelberg Institute for Theoretical Studies (HITS), Schloss-Wolfsbrunnenweg 35, 69118 Heidelberg, Germany\\
$^{3}$Zentrum f\"{u}r Astronomie der Universt\"{a}t Heidelberg, Landessternwarte,  K\"{o}nigstuhl 12, 69117, Heidelberg, Germany\\
$^{4}$Sydney Institute for Astronomy (SIfA), School of Physics, University of Sydney, NSW 2006, Australia\\
$^{5}$Stellar Astrophysics Centre, Department of Physics and Astronomy, Aarhus University, Ny Munkegade 120, DK 8000 Aarhus C, Denmark\\
$^{6}$School of Physics, University of New South Wales, NSW 2052, Australia\\
$^{7}$Institute for Astronomy, University of Hawai`i, 2680 Woodlawn Drive, Honolulu, HI 96822, USA\\
$^{8}$Institut f\"{u}r Astrophysik, Georg-August-Universit\"{a}t G\"{o}ttingen, Friedrich-Hund-Platz 1, 37077 G\"{o}ttingen, Germany\\
$^{9}$Center for Space Science, NYUAD Institute, New York University Abu Dhabi, PO Box 129188, Abu Dhabi, UAE\\
$^{10}$Kapteyn Astronomical Institute, University of Groningen, Groningen, 9700 AV, The Netherlands\\
$^{11}$Department of Astronomy, Beijing Normal University, Beijing 100875, People's Republic of China
}

\date{Accepted XXX. Received YYY; in original form ZZZ}

\pubyear{2015}

\begin{document}
\label{firstpage}
\pagerange{\pageref{firstpage}--\pageref{lastpage}}
\maketitle

% Abstract of the paper
\begin{abstract}
Mass loss by red giants is an important process to understand the final stages of stellar evolution and the chemical enrichment of the interstellar medium. Mass-loss rates are thought to be controlled by pulsation-enhanced dust-driven outflows.
Here we investigate the relationships between mass loss, pulsations, and radiation, using 3213 luminous {\it Kepler} red giants and 135000 ASAS--SN semiregulars and Miras. \redbf{Mass-loss rates are traced by infrared colours using 2MASS and {\it WISE} and by observed-to-model {\it WISE} fluxes, and are also estimated using dust mass-loss rates from literature assuming a typical gas-to-dust mass ratio of 400.}  To specify the pulsations, we extract the period and height of the highest peak in the power spectrum of oscillation. Absolute magnitudes are obtained from the 2MASS $K_s$ band and the Gaia DR2 parallaxes. Our results follow. (i) Substantial mass loss sets in at pulsation periods above $\sim$60 and $\sim$100 days, corresponding to Asymptotic-Giant-Branch stars at the base of the period-luminosity sequences C$'$ and C. (ii) The mass-loss rate starts to rapidly increase in semiregulars for which the luminosity is just above the red-giant-branch tip and gradually plateaus to a level similar to that of Miras. (iii) \redbf{The mass-loss rates in Miras do not depend on luminosity, consistent with pulsation-enhanced dust-driven winds.} (iv) \redbf{The accumulated mass loss on the Red Giant Branch consistent with asteroseismic predictions reduces the masses of red-clump stars by $6.3$\%, less than the typical uncertainty on their asteroseismic masses}. Thus mass loss is currently not a limitation of stellar age estimates for galactic archaeology studies.
\end{abstract}

% \redbf{Mass-loss rates are traced by infrared colours using 2MASS and {\it WISE} ($K_s-$[12] and $K_s-$[22]); these are consistent with observed-to-model
% {\it WISE} fluxes.} 

% Select between one and six entries from the list of approved keywords.
% Don't make up new ones.
\begin{keywords}
Stars: oscillations--Stars: mass-loss--Stars: late-type--Techniques: photometric
\end{keywords}

%%%%%%%%%%%%%%%%%%%%%%%%%%%%%%%%%%%%%%%%%%%%%%%%%%

%%%%%%%%%%%%%%%%% BODY OF PAPER %%%%%%%%%%%%%%%%%%

\section{Introduction}

% \subsection{Mass loss in asymptotic giant branch stars}
Mass loss, pulsations, and radiation are three fundamental processes associated with Asymptotic Giant Branch (AGB) stars. Mass loss in AGB stars is generally considered to be caused by pulsation-enhanced dust-driven outflows \citep[see][for a recent review]{hofner2018a}. In this scenario, pulsations levitate the upper atmospheres of the star to higher radii, with \redbf{low gas temperatures} of about 1000 K. These conditions favour the formation of dust grains, such as Mg/Fe silicates in oxygen-rich AGB stars and amorphous carbon in carbon-rich AGB stars. The dust grains in the circumstellar envelope are accelerated outwards by the radiation pressure from the star, and \redbf{are pushing the gas} to which they are collisionally coupled, causing mass loss. Although the pulsation-enhanced dust-driven model is widely accepted as a reasonable description of mass loss in AGB stars, the underlying details are still poorly understood. 

Mass loss and pulsations were initially linked for AGB stars having large amplitude and long-period pulsations \citep[$\gtrsim$ 300 days, e.g.,][]{cannon1967a, habing1996a}. These studies were later extended to lower-luminosity red giants pulsating in a shorter period regime (\mbox{$\lesssim$ 300 days}), which is also the focus of our study. \citet{glass2009a} and \citet{mcdonald2016a} found a critical pulsation period of 60 days, above which a rapid and systematic increase of dust production sets in. This period is just above the upper limit for the Red Giant Branch (RGB) tip of 20--50 days \citep[e.g.,][]{kiss2003a}, meaning that the stars are just above the RGB tip in the H--R diagram. This critical period approximately corresponds to the period at the base of period--luminosity (P--L) sequence C$^{\prime}$ (e.g., \citealt{mcdonald2019a}; sequences A, B, C$^{\prime}$, C and D are labelled according to the nomenclature used by \citealt{wood2015a} and are shown in Fig.~\ref{fig:massloss}a). \citet{riebel2012a} and \citet{mcdonald2019a} found that stars on sequences C$^{\prime}$ and C generally have higher mass-loss rates (\mbox{10$^{-8}$--10$^{-7}$\msun/yr}) than stars on sequences B and A (10$^{-10}$--10$^{-9}$\msun/yr). Analogous to stars on sequence C$^{\prime}$, the similar and high mass-loss rates of stars on sequence C suggest that there could be another rapid and systematic increase of dust production at $\sim$100 days, which roughly corresponds to the base of sequence C.

A strong positive correlation between the mass-loss rate and luminosity in long-period AGB stars (\mbox{$\gtrsim$ 300 days}) has been recently reported by \citet{danilovich2015a} and  \citet{groenewegen2020a}. This suggests that the effect of radiation on dust is effective in driving superwinds in these stars. Does this correlation exist in short-period AGB stars (\mbox{$\lesssim$ 300 days})?  An interesting case would be that of metal-poor oxygen-rich stars, where the low metallicity makes it difficult for a considerable amount of \redbf{dust to form in the atmosphere due to a lack of condensable material}. As a result, if mass-loss is indeed driven by radiation pressure, metal-poor stars should have lower mass-loss rates. However, \redbf{dust production is commonly observed in metal-poor environments}, including several globular clusters \citep{lebzelter2006a, van-loon2006b, boyer2009a, boyer2010a, mcdonald2011c, mcdonald2011a, origlia2014a}. Furthermore, \citet{mcdonald2016a} analysed a small sample of nearby stars using \textit{Hipparcos} parallaxes and found tentative evidence that \redbf{mass loss is independent of stellar luminosity in short-period (\mbox{$\lesssim$ 300 days}) stars (see their Fig. 2a)}. Therefore, it is of prime importance to understand whether radiation drives appreciable mass loss in short-period AGB stars (\mbox{$\lesssim$ 300 days}).

 Understanding mass loss has important implications for RGB stars, in particular for determining ages of red clump stars that are used extensively as probes for galactic archaeology \citep[e.g.,][]{girardi2016a,silva-aguirre2018a}. \citet{groenewegen2014a} made the first detection of rotational CO line emission in an RGB star, which was interpreted as direct evidence of mass loss. If appreciable mass is lost in the RGB phase, the primordial masses of red-clump stars will be underestimated, which will lead to an overestimation in their ages. \citet{miglio2012a} and \citet{kallinger2018a} performed asteroseismic analysis of a few tens of cluster red giants and found an integrated mass loss of $\sim$0.08 \msun\ on the RGB (or 7.3\%). The \kep\ mission has observed a few thousand luminous red giants \citep{banyai2013a, mosser2013a, stello2014a, yu2020a}, and their high-precision data have motivated us to explore mass loss in RGB stars as an ensemble. 

In this work, first of all, we search for critical pulsation periods above which substantial mass loss sets, in using luminous \kep\ red giants \citep{banyai2013a, mosser2013a, stello2014a, yu2020a}, as well as \mbox{ASAS--SN} Long Period Variables (LPVs) \citep{jayasinghe2019b}. Our sample is composed of 3213 \kep\ LPVs taken from \citet[][hereafter Paper I]{yu2020a}, supplemented by 134,928 \mbox{ASAS--SN} LPVs \citep{shappee2014a, kochanek2017a} with light-curve duty cycles \footnote{We define the duty cycle as the fraction of data cadences within the span of observations that contain valid data. Ground-based one-site observations generally have relatively low duty cycles due to  the diurnal cycle.} better than 0.25. \kep\ \mbox{long-cadence} data are well suited for this investigation because pulsation periods can be precisely measured from light curves with a sampling interval (29.4 minutes) covering four years. The \kep\ sample is also dominated by stars on the RGB, and thus is valuable for exploring mass loss in this phase. The ASAS--SN survey has observed hundred of thousands of LPVs, and can add significantly to the limited number of Miras in the \kep\ sample. Second, Gaia DR2 parallaxes can be used to estimate the luminosities for the LPVs, providing an excellent opportunity to explore radiation effects on mass loss. Finally, we aim to estimate the typical integrated mass loss on the RGB, which enables us to understand the effect of the integrated mass loss on the ages of red-clump stars. 

\begin{figure*}
\begin{center}
\resizebox{0.33\textwidth}{!}{\includegraphics{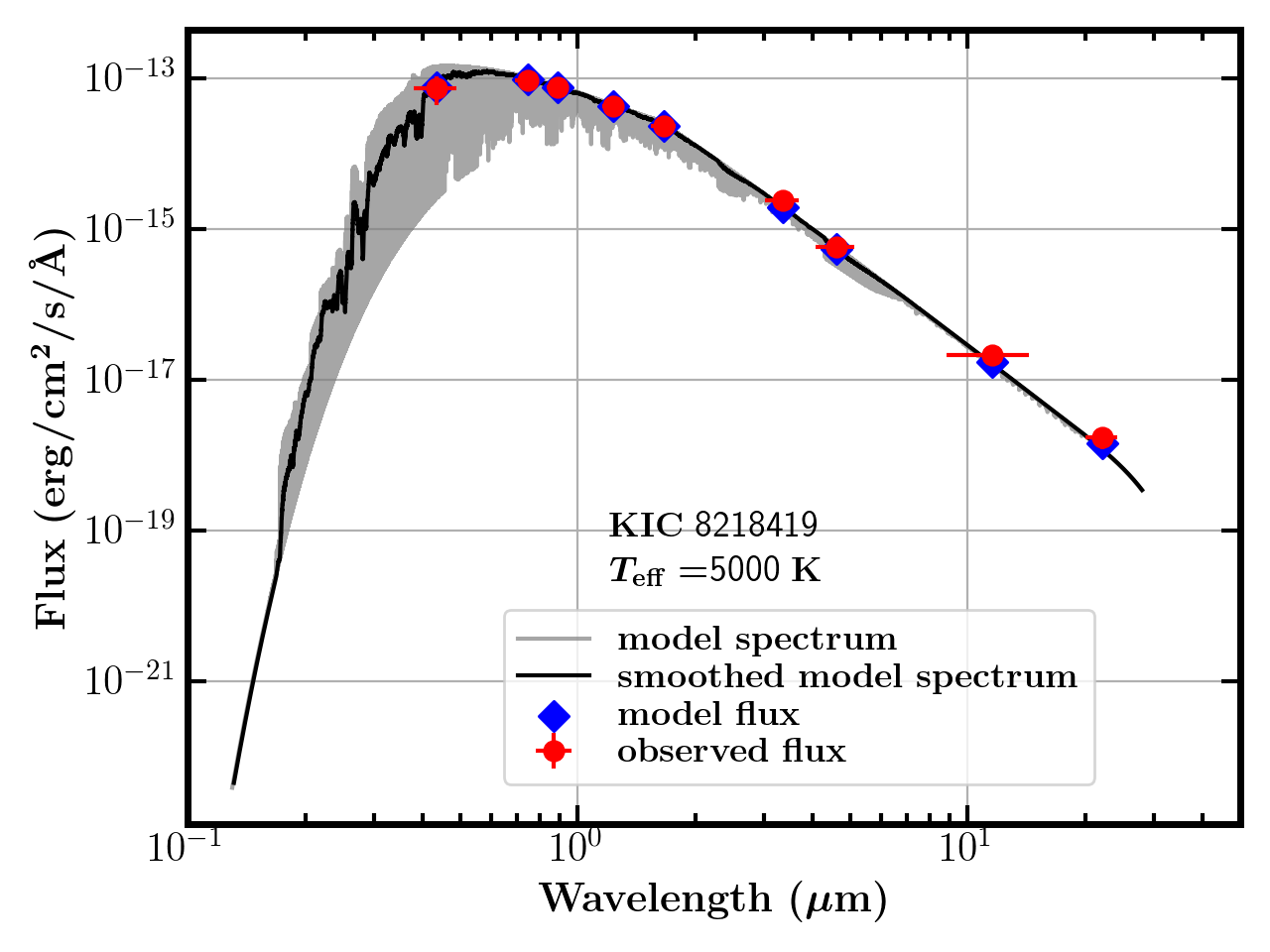}}
\resizebox{0.33\textwidth}{!}{\includegraphics{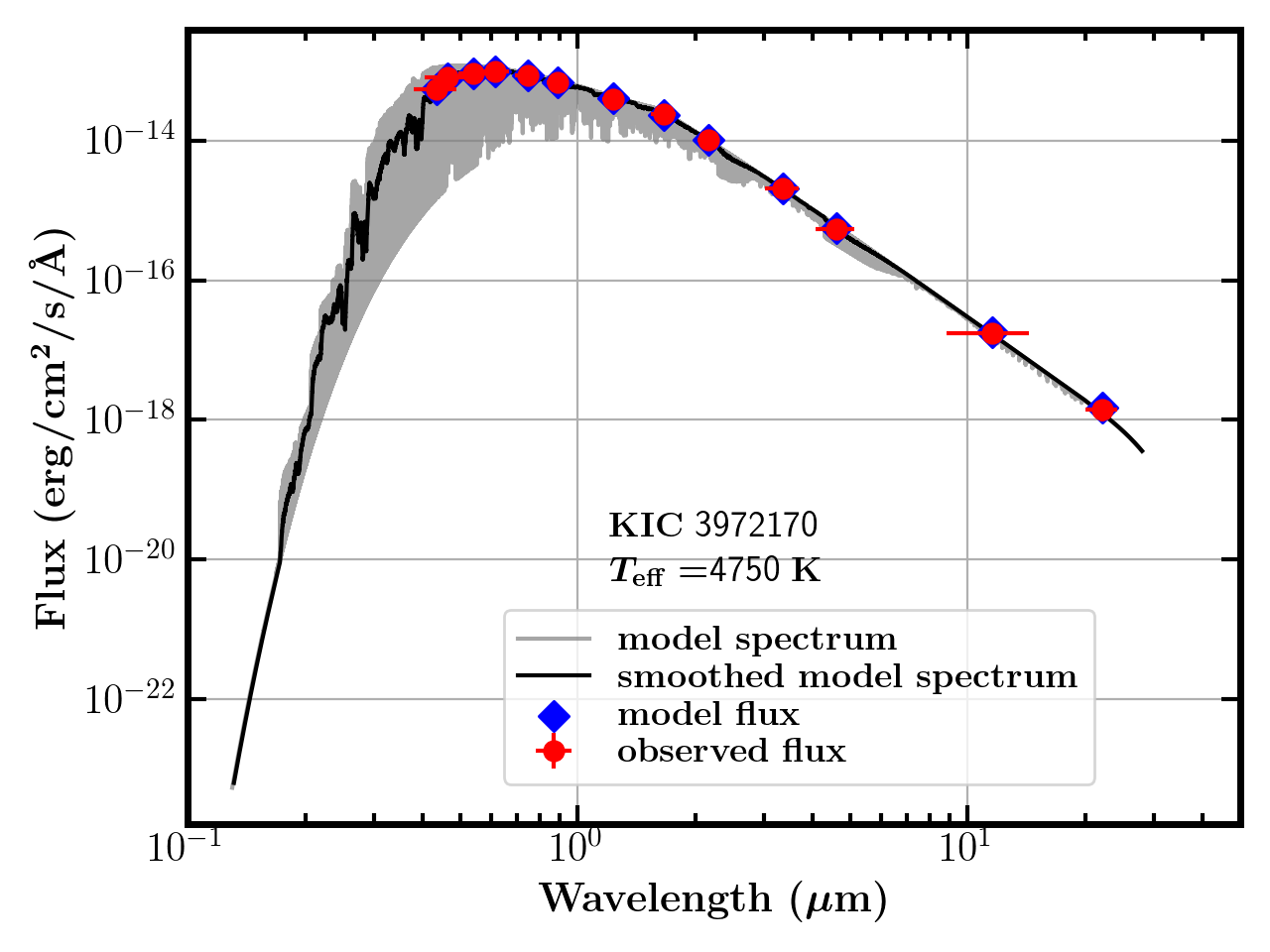}}
\resizebox{0.33\textwidth}{!}{\includegraphics{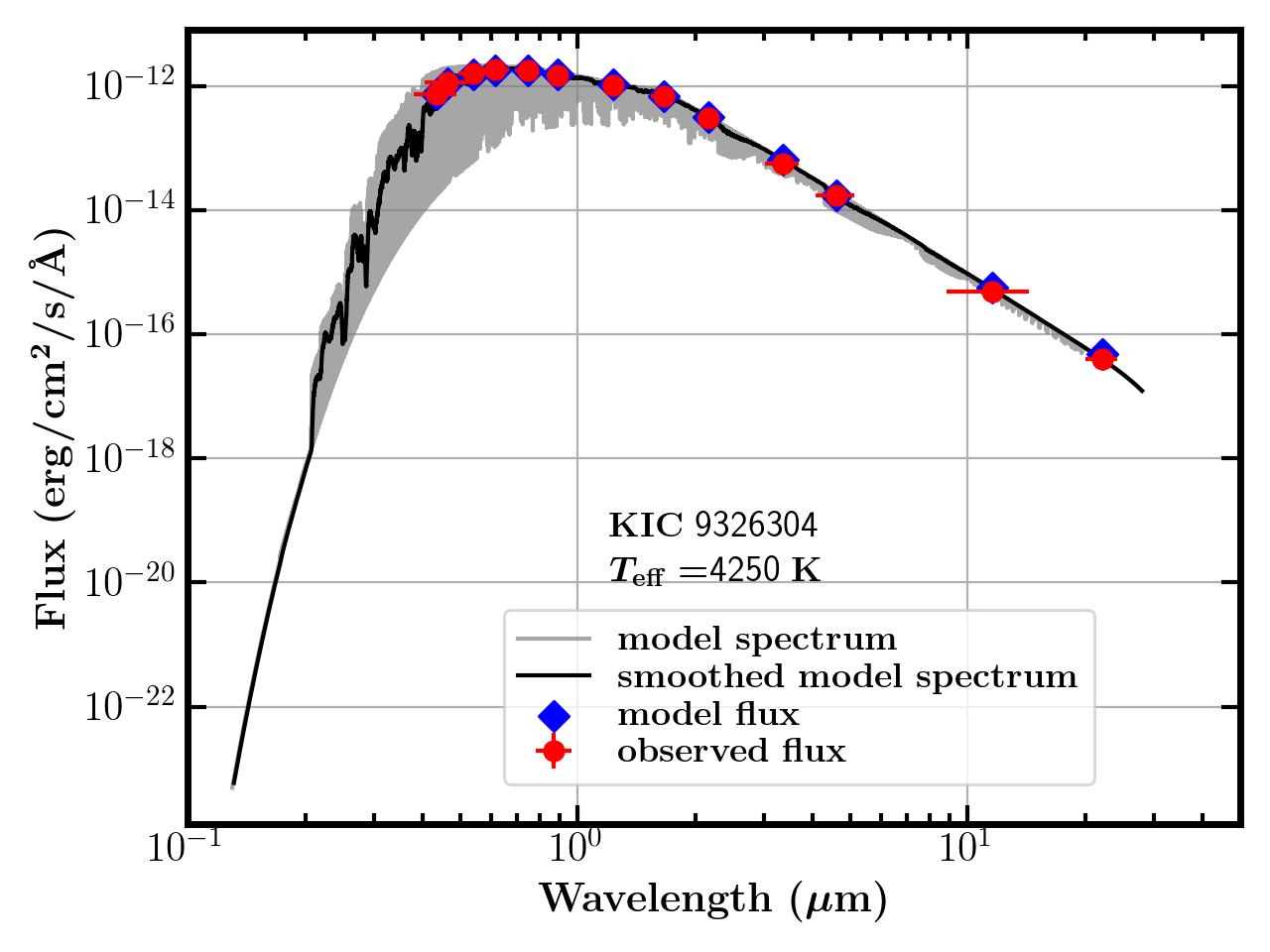}}\\
\resizebox{0.33\textwidth}{!}{\includegraphics{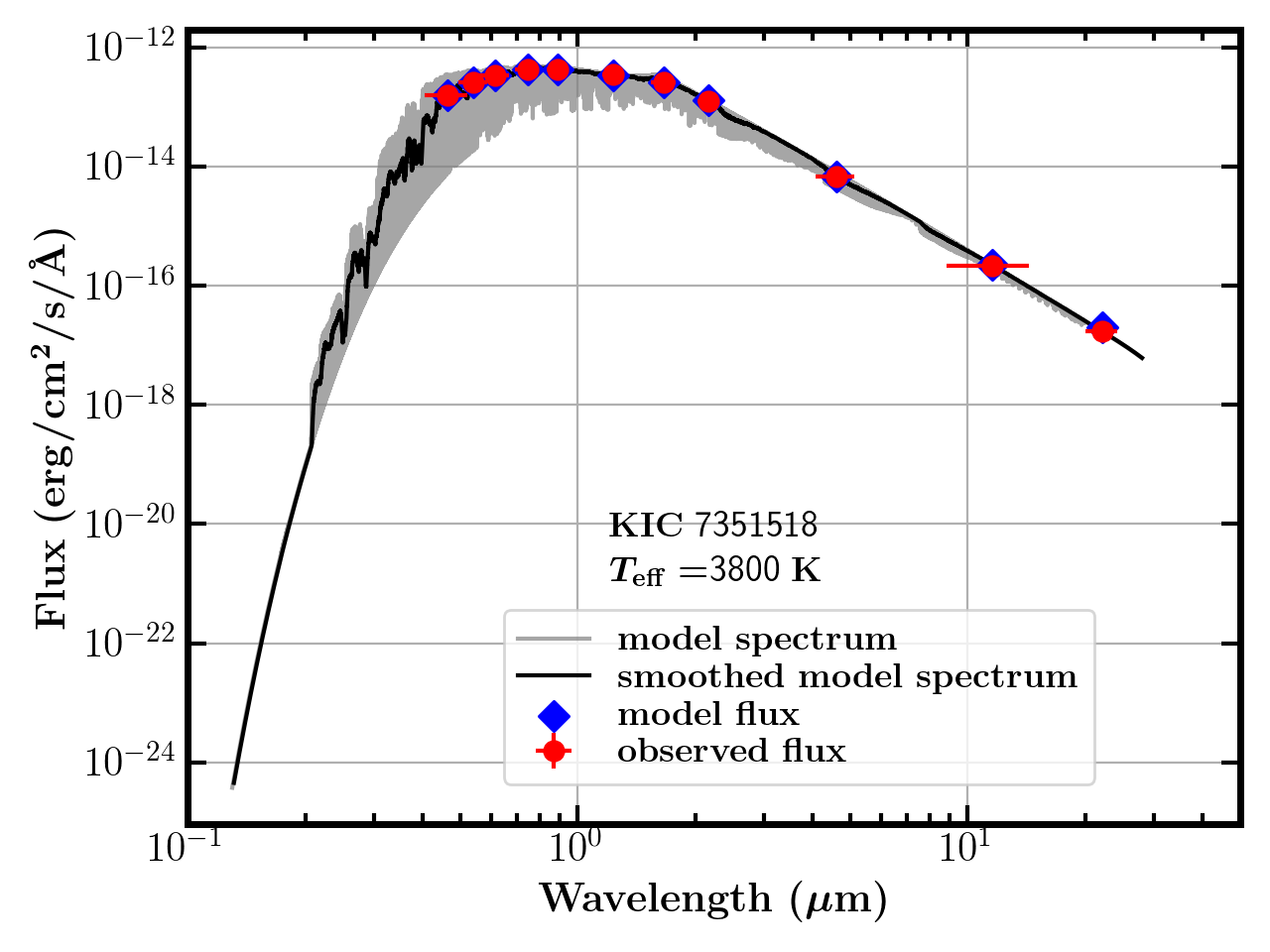}}
\resizebox{0.33\textwidth}{!}{\includegraphics{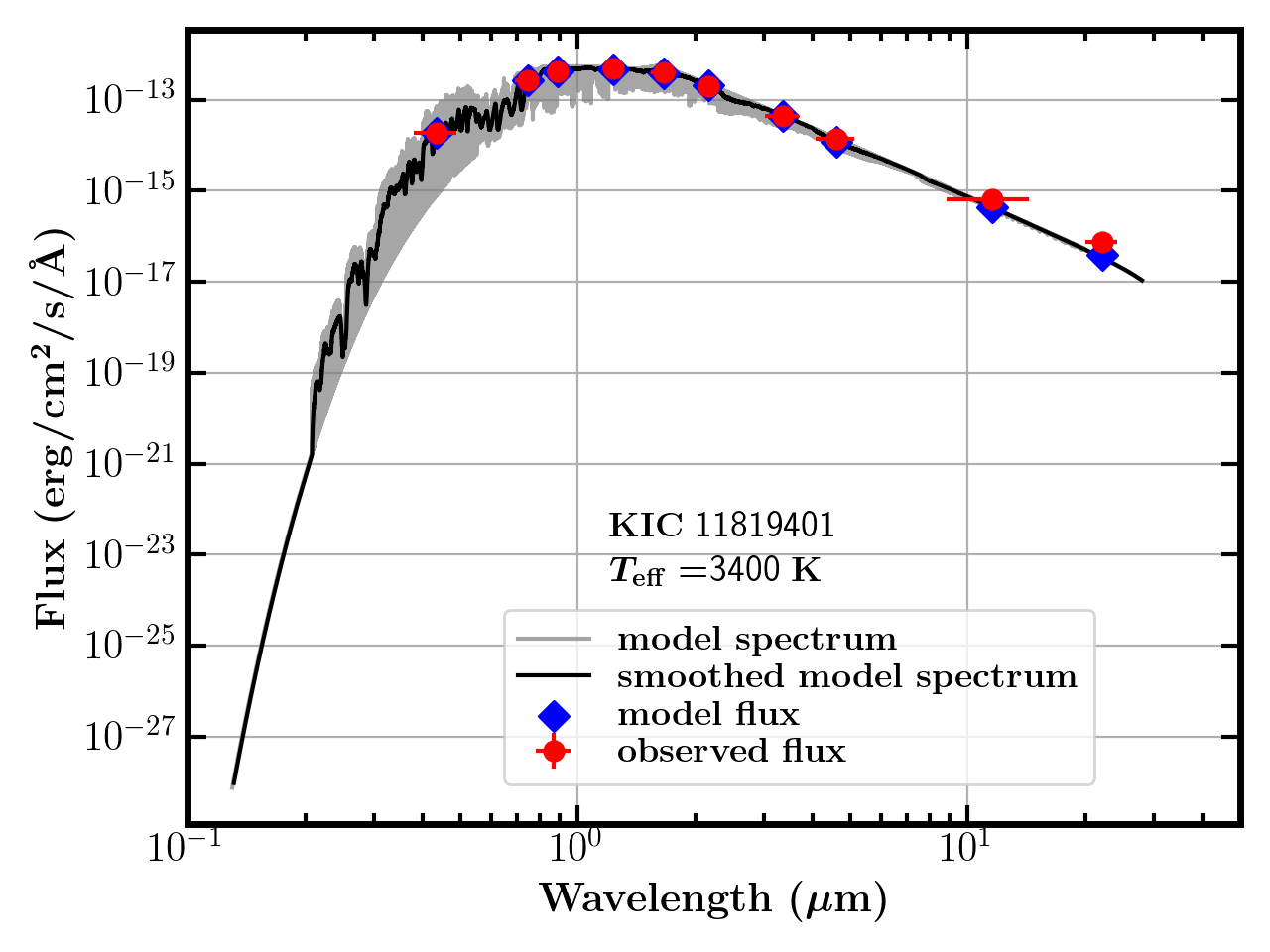}}
\resizebox{0.33\textwidth}{!}{\includegraphics{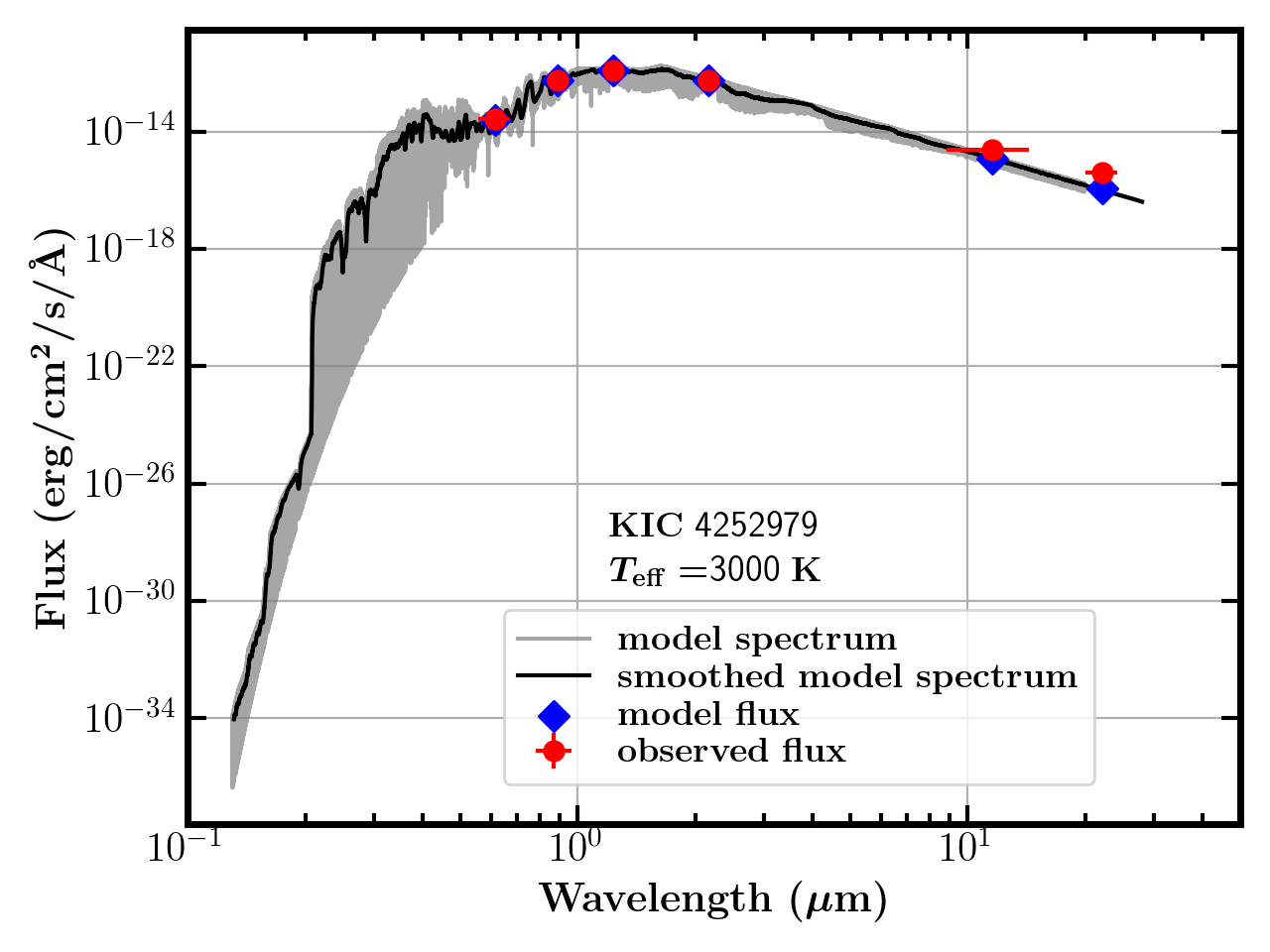}}\\
\caption{Six representative SED fits.  Effective temperatures of the best-fitting MARCS spectra range from 5000 K down to 3000 K. In each panel, the MARCS spectrum is shown in grey and its smoothed spectrum (only for display, not used for the SED fitting) is shown in black. The model and observed flux densities are indicated in the blue diamonds and red circles, respectively, with error bars indicated wherever available.}
\label{fig:sedfits}
\end{center}
\end{figure*}

\section{Data and Methodology}\label{datamethod}
\subsection{Data}
We revisited the sample of 3213 \kep\ LPVs from \citetalias{yu2020a}, which includes Miras, semiregulars (SRs), and red giants of lower luminosity. The pulsation period and amplitude of the dominant mode (the highest peak of in the Fourier spectrum) for each star are also taken from \citetalias{yu2020a}. The periods of the dominant modes are all longer than 1 day. Radial orders of the dominant modes range from $n=1$ to $n=6$.

In addition to the \kep\ sample, we included 134,928 \mbox{ASAS--SN} LPVs \citep{jayasinghe2019b} to enlarge the sample of Miras and SRs. These LPVs were selected to have a  duty cycle higher than 0.25, in order to obtain  robust measurements of pulsation periods and amplitudes. We retrieved pulsation periods and amplitudes from the ASAS--SN light  curves \citep{shappee2014a, kochanek2017a}. The power spectra of the light curves were calculated using an oversampling factor of 15 to ensure that the dominant modes were correctly identified. For each star, we searched the power spectrum for the highest peak and estimated the pulsation amplitude by its height. For the ASAS--SN LPVs, the absolute magnitudes \mk\ were derived using distances from \citet{bailer-jones2018a} based on Gaia DR2 parallaxes \citep{lindegren2018a} and using extinctions from the ASAS--SN catalogue \citep{jayasinghe2019b}. The extinctions therein were based on the total \mbox{line-of-sight} Galactic reddening E(B--V) from the recalibrated SFD dust maps \citep{schlegel1998a,schlafly2011a}. For the \kep\ sample \mk\  was derived in a similar way, and we refer the reader to \citetalias{yu2020a} for more details.

\subsection{Methodology}
% \redbf{We used infrared colors from 2MASS and WISE photometry and observed-to-model infrared flux ratios determined through spectral energy distribution (SED) fitting to trace total (gas+dust) mass-loss rates\footnote{When mass-loss rates are referred to, we mean total mass-loss rates, unless either gas or dust mass-loss rates are specifically mentioned otherwise}. We also estimated total mass-loss rates using dust mass-loss rates from \citet{riebel2012a} that were based on a \mbox{grid-based} modelling approach. Each of these three methods are detailed below.}

\redbf{In this section we will describe three independent methods used to study mass loss in LPVs: (1) Tracing mass-loss rates \footnote{When mass-loss rates are referred to, we mean total (gas+dust) mass-loss rates, unless otherwise stated.} using infrared colours from 2MASS and {\it WISE} photometry, (2) tracing mass-loss rates by comparing flux ratios between observed and modelled spectral energy distributions (SED), and (3) estimating total mass loss rates using dust mass-loss rates from \citet{riebel2012a} using grid-based modelling.}

\subsubsection{Mass-loss rate tracer: infrared colours}
\redbf{The \kwfour\ colour\footnote{Here [22] stands for the magnitude of the W4 band (22.09 $\mu$m) of the Wide-field Infrared Survey Explorer (\textit{WISE}) space observatory. Similarly, [12] stands for the magnitude of the \textit{WISE} W3 band (11.56 $\mu$m).} is known to be an approximate measure of mass-loss rates, to within an order of magnitude \citep{mcdonald2016a,mcdonald2018b}. We note that the mass-loss rates span at least four orders of magnitude for the stars in the sample. The underlying principle of using the colour as a mass-loss tracer is that the \kwfour\ colour traces dust column density, and thus the dust mass-loss rate, by assuming a  typical wind velocity (10 km/s, as used by \citealt{srinivasan2011a}). This dust mass-loss rate can then be multiplied by a typical gas-to-dust mass ratio (400, as used by \citealt{mcdonald2019a}) to estimate the total mass-loss rate.} 

\redbf{We should keep in mind the distinction between infrared colours, which are presumably dominated by dust, and total mass-loss rates, which are dominated by the gas. Although it is reasonable to convert the infrared colour to total mass-loss rates assuming a fixed wind velocity and a fixed gas-to-dust mass ratio, a more accurate way of estimating mass-loss rates uses observations of CO (or other molecules) emission lines in the sub-mm regime, which makes it possible to directly determine gas densities in the circumstellar envelope and the wind velocity \citep[e.g.,][]{mcdonald2018b, diaz-luis2019a}. A drawback of this CO-emission-line method, however, is that it can only be applied to a small sample of LPVs due to observational efficiency, far less than the size of our sample. Thus, without converting the \kwfour\ colour to the total mass-loss rate in this work, we used the colour in this work to investigate the ensemble property of mass-loss rates and the dependence of mass-loss rates on pulsations and radiation. Furthermore, we also used the \kwthree\ colour to trace mass-loss rates, considering the higher photometry sensitivity in the W3 band compared to W4 \citep[e.g.,][]{mcdonald2019b}.}

\subsubsection{Mass-loss rate tracer: observed-to-model infrared fluxes}\label{sed}
In addition to using \kwthree\ and \kwfour\ colours to estimate mass-loss rate, we also used the ratio of the observed to predicted flux densities of the \textit{WISE} W3 and W4 bands. For this, we built for each \kep\ star an observed SED and its best-fitting model SED. The predicted flux densities were obtained from the best-fitting MARCS spectra \citep{gustafsson2008a} by carrying out SED fitting.

\begin{figure}
\begin{center}
\resizebox{\columnwidth}{!}{\includegraphics{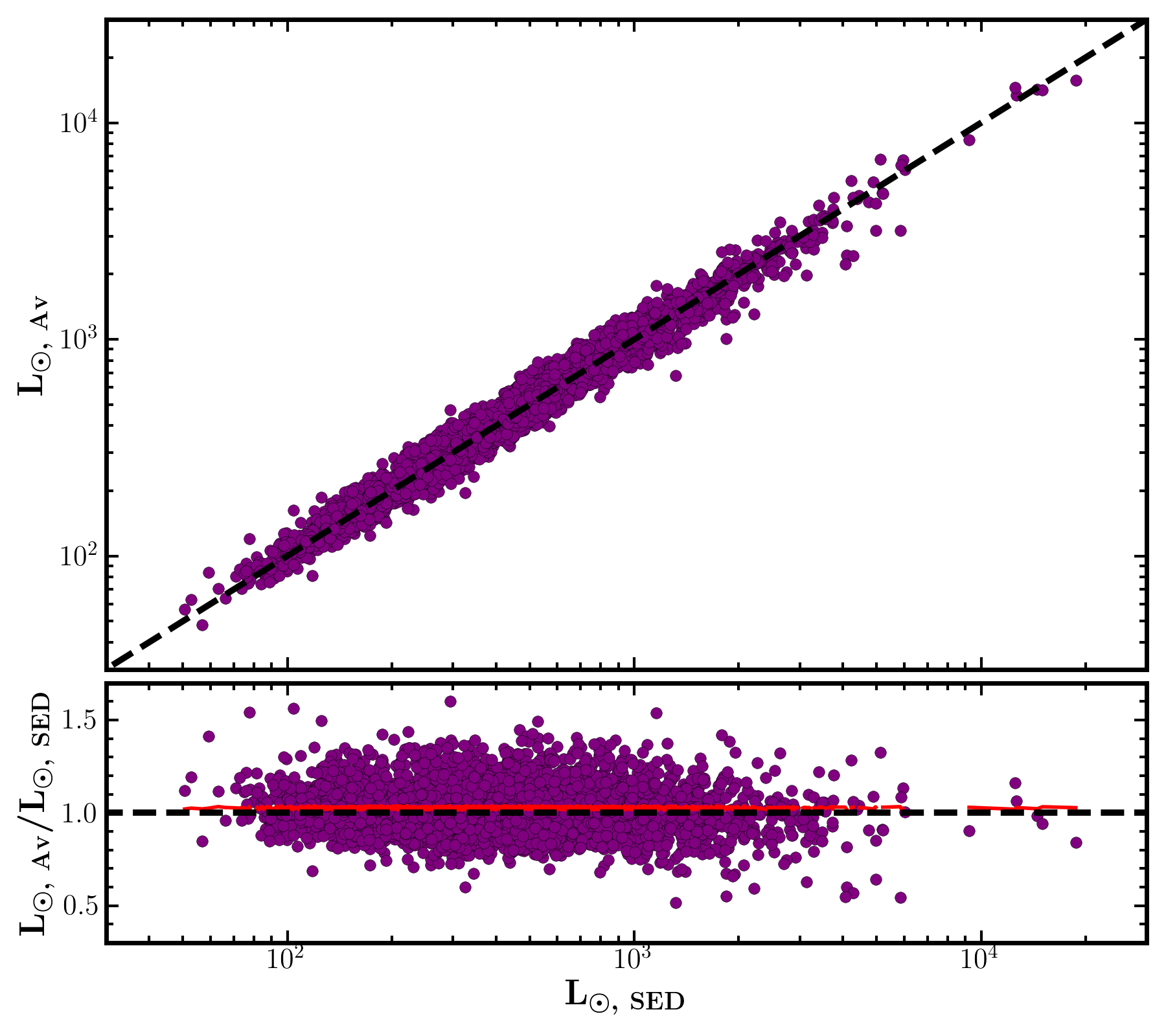}}\\
\caption{Comparison of luminosities derived from SED-fitted bolometric fluxes and distances, namely $\rm L_{\odot, SED}$, and  luminosities computed from 2MASS $K_{\rm{s}}$ magnitudes and distances with extinction and bolometric corrections, namely $\rm L_{\odot, Av}$. The running mean of the luminosity ratio $\rm L_{\odot, Av}$/$\rm L_{\odot, SED}$ is shown in red in the lower panel.} 
\label{fig:lumicomp}
\end{center}
\end{figure}

To construct an observed SED, we combined broadband photometric data wherever available using: (1) $B_T$ (0.428 $\mu$m) and  $V_T$ (0.534 $\mu$m) magnitudes from the {\it Tycho2} catalogue \citep{hog2000a}; (2) B (0.433 $\mu$m) and V (0.540 $\mu$m) magnitudes from the UCAC4 \citep{zacharias2012a} and APASS \citep{henden2016a} catalogues; (3) $g$ (0.464 $\mu$m), $r$ (0.612 $\mu$m), $i$ (0.744 $\mu$m), and $z$ (0.890 $\mu$m) magnitudes from the \kep\ Input Catalogue \citep{brown2011a}; (4) $G_{\rm BP}$ (0.528 $\mu$m) and $G_{\rm RP}$ (0.788 $\mu$m) magnitudes from Gaia DR2 \citep{evans2018a}; (5) $J$ (1.23 $\mu$m), $H$ (1.66 $\mu$m), and $K_{\rm s}$ (2.16 $\mu$m) magnitudes from the 2MASS survey \citep{cutri2003a}; (6) W1 (3.35 $\mu$m), W2 (4.60 $\mu$m), W3 (11.56 $\mu$m), and W4 (20.09 $\mu$m) magnitudes from the AllWISE catalogue \citep{cutri2013a}. Only high quality photometric data were retained for further analysis by setting the most stringent quality flags for each catalogue.  \redbf{Generally, the stars in our sample were observed multiple times by the individual photometric surveys, and the magnitudes used in our work are mean values, which helps to reduce the scatter due to stellar variability.} All magnitudes were converted to flux densities, $F_\lambda$, in units of $\rm erg/cm^2/s/\mathring{\rm A}$ using zero-point flux densities archived by SVO\footnote{http://svo2.cab.inta-csic.es/svo/theory/fps/}.

\begin{figure*}
\begin{center}
\resizebox{0.7\textwidth}{!}{\includegraphics{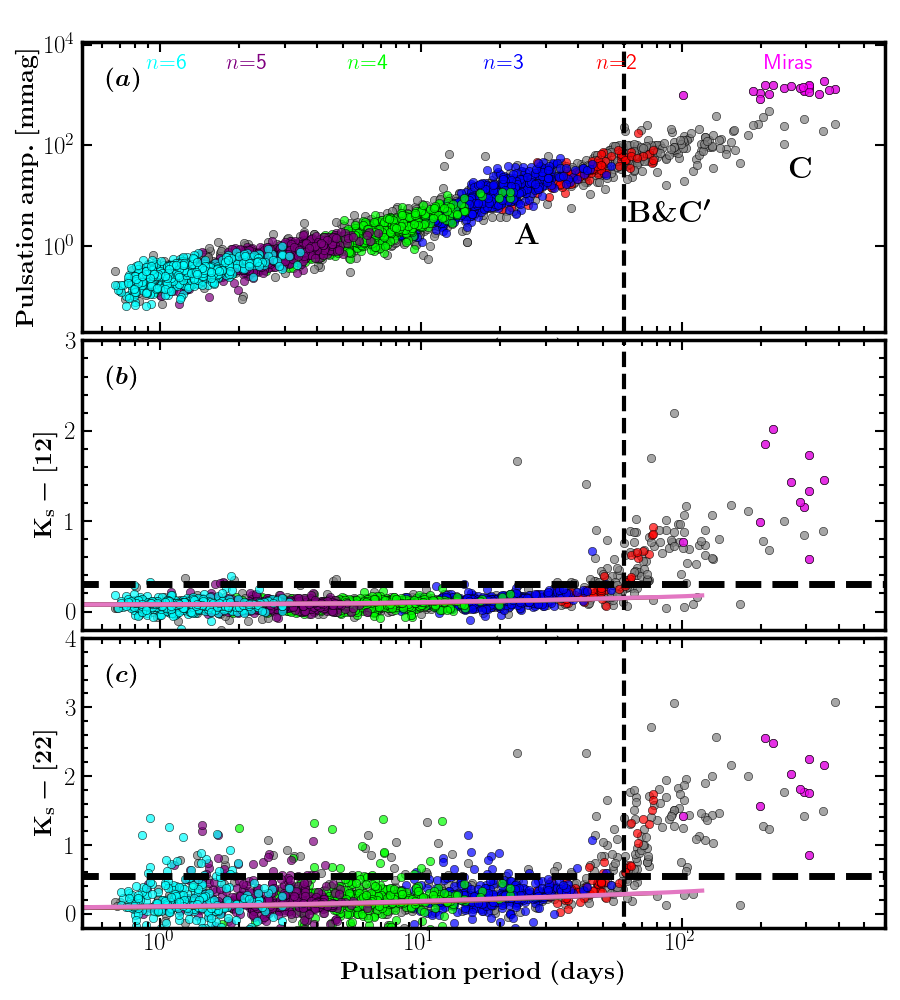}}\\
\caption{\textbf{(a)} Relation between the pulsation amplitude and period of the dominant mode for the sample of 3213 \kep\ LPVs from \citetalias{yu2020a}. Radial orders of the dominant modes are indicated by colour (see labels at the top), except for the stars whose radial orders are unidentified due to short light curve coverage. This assignment to the $n$ sequences shown is as per \citetalias{yu2020a}. The approximate periods of the P--L sequences A, B, C$^{\prime}$, and C are also indicated. \textbf{(b)} Infrared colour \kwthree\ as a function of pulsation period. The pink line depicts the 0.8\msun\ dust-free model as shown in Fig.~\ref{fig:hrmodel}b. Note that all seven models shown in Fig.~\ref{fig:hrmodel}b are plotted but they overlap and only the 0.8\msun\ model is visible, because of negligible colour difference at a given period. The vertical dashed line marks the approximate period threshold, 60 days, of substantial mass loss, while the horizontal dashed line indicate the criterion, \kwthree >0.3 mag to define stars with dust production. \textbf{(c)} Same as panel \textbf{b} except for using \kwfour. The horizontal dashed line indicate the criterion, \kwfour\ >0.55 mag, equivalent to \kwthree>0.3 mag.} 
\label{fig:excesskepler}
\end{center}
\end{figure*}

To build a model SED, we used MARCS spectra \citep{gustafsson2008a}, which are based on a grid of one-dimensional, hydrostatic, plane-parallel (for dwarfs and subgiants) and spherical (for giants) LTE model atmospheres \citep{gustafsson2008a}. The effective temperatures of the models range from 2500 to 4000 K in steps of 100 K and from 4000 to 8000 K in steps of 250 K. The surface gravities vary between $-$1.0 and 5.5 in steps of 0.5 dex. These limits in effective temperature and surface gravity cover the parameter range of the LPVs in the sample. MARCS synthetic spectra were created with a spectral resolution R = 20,000 over the range 0.13--20$\mu$m, and were parametrised by \teff, \logg, [Fe/H], mass, and microturbulence velocity $\nu_{\rm micro}$.

In this work, we left \teff\ and \logg\ as free parameters to search for best-fitting model spectra, and fixed [Fe/H]=0, $M$=1.0 \msun, and $\nu_{\rm micro}$ = 2 km/s. Fixing these parameters is justified by the fact that for our study the best-fitting model spectra depend weakly on [Fe/H], mass, and $\nu_{\rm micro}$ \citep{mcdonald2009a}. These values for [Fe/H], $M$, and $\nu_{\rm micro}$ were selected to represent the ensemble properties of the sample stars, as well as to ensure that we have the best coverage of MARCS atmosphere models at different \teff\ and \logg. Considering the maximum wavelength of the MARCS models is 20 $\mu$m, whereas the \textit{WISE} W4 band extends to $\sim$28 $\mu$m, we extrapolated the spectra beyond 20 $\mu$m using a third-order polynomial fit to the logarithm of the fluxes within the wavelength range 10–20 $\mu$m.

MARCS spectra were convolved with the filter transmission of each adopted band to obtain the flux density. Those spectra with \teff\ and \logg\ that were consistent with the observed counterparts to within 300 K and 1.0 dex, respectively, were used in the following analysis. We adopted the estimates of \teff\ and \logg\ from \citet{mathur2017a}. In addition to \teff\ and \logg, extinction $A_V$ and scale factor $f$ (related to the luminosity and distance of the target star) were allowed to vary freely. The extinction law by \citet{wang2019a} was used for the Gaia $G_{\rm BP}$ and $G_{\rm RP}$ bands and that by \citet{cardelli1989a} was used for the other bands.

The commonly used $\chi^2$ statistic was not suitable for searching for best-fitting models, given that magnitude uncertainties were unavailable for some bands. Instead, we minimised %a pseudo-$\chi^2$ statistic: 
\begin{equation}
\sum_{i}\frac{1}{F_{\lambda,i,\rm{obs}}^2}
\left(F_{\lambda,i, \rm{obs}}-F_{\lambda,i, \rm{mod}}\right)^2,
\end{equation}
where $F_{\lambda,i, \rm{obs}}$ and $F_{\lambda,i, \rm{mod}}$ are the observed and model flux densities in the $i^{\rm th}$ band, respectively. Thus, we minimised the sum of the squared relative differences. This was to ensure that the SED fit was not dominated by a few bands with high fluxes, for which the differences between $F_{\lambda,i, \rm{obs}}$ and $F_{\lambda,i, \rm{mod}}$ dominate. The \textit{WISE} W3 and W4 bands were not used for the minimisation because they could bias the SED fitting when substantial flux excess is observed. 

In the course of the minimisation, we repeated the above procedures until the largest difference between the observed and predicted flux densities across all bands was less than 10\%. If the largest difference was greater than 10\%, the associated data point was considered to be an outlier and hereafter clipped. The clipping step was applied to discard poor photometry, which generally removed a few data points (bands) for each star. This step was necessary because while stringent quality flags were set to attempt to select good photometry, poor data could also be retained and misused to build an observed SED \citep[e.g., see][]{dennihy2020a}. We found that this step significantly improved the SED fitting.

Fig.~\ref{fig:sedfits} shows six representative fits with effective temperatures of the best-fitting model spectra ranging from 5000~K down to 3000~K. The SED fits show a good match between observed (red circles) and calculated (blue diamonds) flux densities in a wide wavelength range $0.4~\mu \rm{m}< \lambda < 22~\mu \rm{m}$. To validate further our SED fitting, we compare in Fig.~\ref{fig:lumicomp} the luminosities derived from two methods: one from SED-fitted bolometric fluxes and distances, namely $\rm L_{\odot, SED}$, and the other one from 2MASS $K_{\rm{s}}$ magnitudes and distances with extinction and bolometric corrections, namely $\rm L_{\odot, Av}$. We refer the reader to \citetalias{yu2020a} for more details on the derivation of $\rm L_{\odot, Av}$. The two sets of luminosites, $\rm L_{\odot, SED}$ and $\rm L_{\odot, Av}$, are consistent well,  with a small systematic offset of 2.6\% and a scatter of 14.1\%.

\begin{figure}
\begin{center}
\resizebox{\columnwidth}{!}{\includegraphics{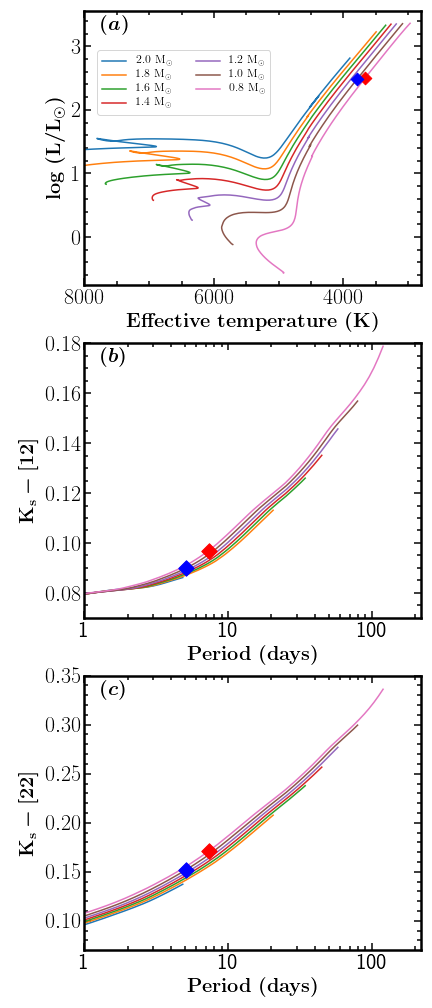}}\\
\caption{(\textbf{a}) Evolutionary tracks for different masses at solar metallicity from the ZAMS to the RGB tip, adopted from \redbf{the MIST database} \citep{dotter2016a, choi2016a}. Two models on the 0.8 and 1.0 $\rm M_{\odot}$ tracks with the same luminosity, \logl=2.5, are highlighted with diamonds. Panels (\textbf{b}) and (\textbf{c}) show dust-free infrared colours \kwthree\ and \kwfour\ as a function of pulsation period, respectively. The synthetic photometry, namely $K_{\rm s}$, [12], and [22], were taken from \redbf{the MIST database}. The periods were computed from the \numax\ scaling relation using \logg\ and \teff\ from the evolutionary tracks shown in panel \textbf{(a)}. The blue and red diamonds in panels \textbf{(b)} and \textbf{(c)} correspond to the two models highlighted in panel \textbf{(a)}.}
\label{fig:hrmodel}
\end{center}
\end{figure}

\begin{figure*}
\begin{center}
\resizebox{0.7\textwidth}{!}{\includegraphics{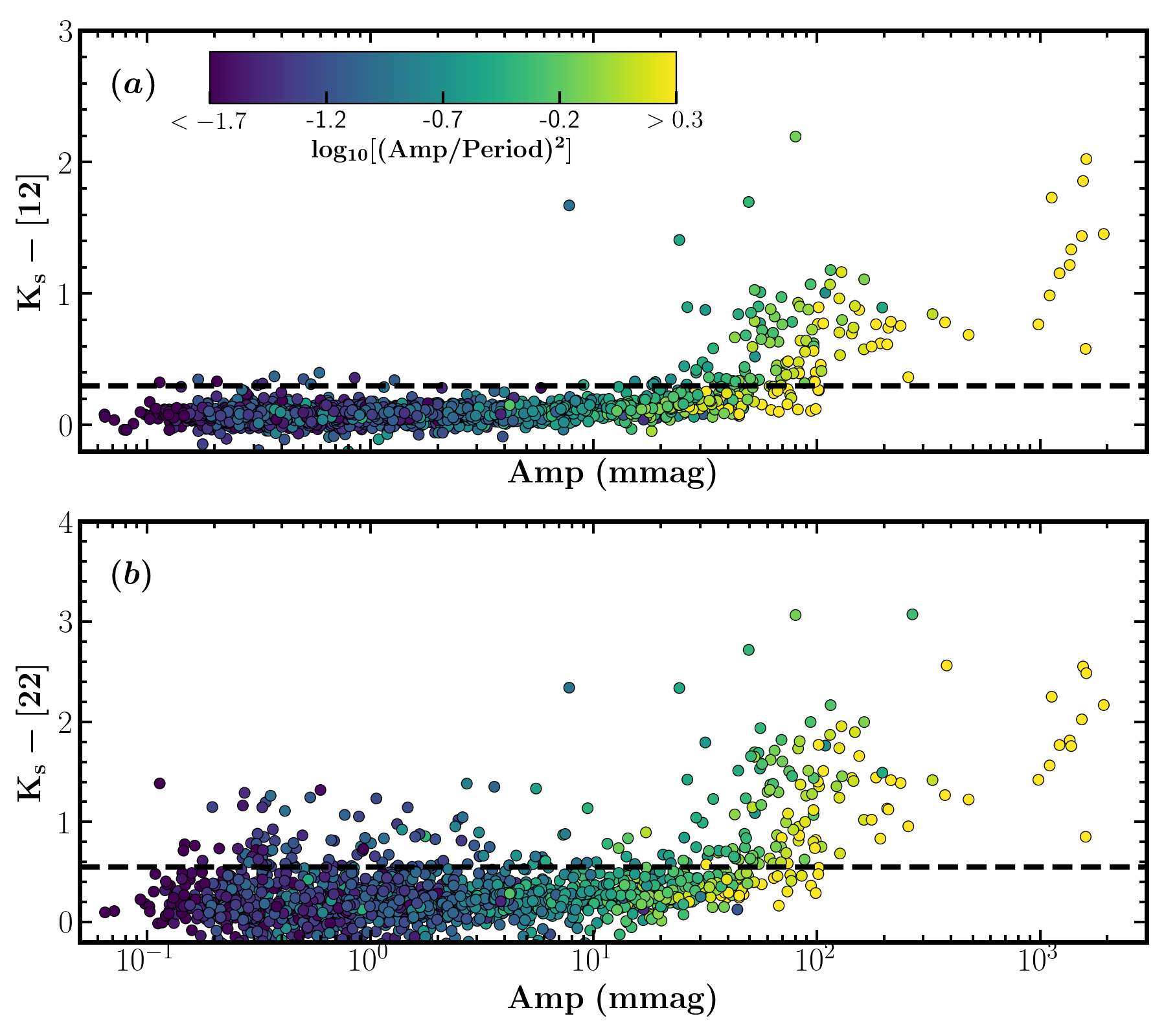}}\\
\caption{\textbf{(a)} \kwthree\ as a function of pulsation amplitude, colour-coded by the logarithmic of the ratio of pulsation amplitude to period squared, the argument of which, i.e., $\rm{(amplitude/period)^2}$, is proportional to the power delivered by a sound wave. The horizontal dashed line indicate the criterion, \kwthree >0.3 mag, or equivalently \kwfour\ >0.55 mag, to define stars with dust production. \textbf{(b)} Same as panel \textbf{a} now using the mass loss indicator \kwfour.} 
\label{fig:ampperiodpower}
\end{center}
\end{figure*}

\subsubsection{Estimating mass-loss rates: grid-based  modelling}\label{gridmodels}
\redbf{To confirm the mass-loss property revealed by the infrared colours and, more importantly, to estimate integrated mass-loss rates of RGB stars, we also estimated the total mass-loss rates for the sample of LPVs in the LMC studied by \citet{riebel2012a}. Those authors obtained dust mass-loss rates by fitting the GRAMS model grid, which is a pre-computed grid of radiative transfer models of evolved stars and circumstellar dust shells composed of either silicate or carbonaceous dust, to multiple bands of photometry from the optical to the mid-infrared. A constant expansion velocity of 10 km/s was assumed therein.} 

\redbf{Similar to \citet{mcdonald2019b}, we converted dust mass-loss rates to total mass-loss rates by multiplying the former by a typical gas-to-dust ratio of 400. We note that a value of 200 has also been widely used for AGB stars in our Galaxy \citep[e.g.,][]{groenewegen2009a,gullieuszik2012a,riebel2012a}. Our main purpose, however, was to evaluate typical total mass loss rates for stars on sequences A and B, in particular those stars near the RGB tip, because these mass loss rates are useful for estimating integrated mass loss of RGB stars. These stars are predominantly oxygen-rich rather than carbon-rich, and thus higher gas-to-dust mass ratios are expected \citep[e.g., 500 adopted for O-rich AGB stars and 200 for C-rich AGB stars by][]{riebel2012a}. Therefore, we adopted a typical gas-to-dust ratio of 400, similar to \citet{mcdonald2019b}. The total mass-loss rates were used to predict integrated mass loss of RGB stars, which we found to be consistent with those obtained from asteroseismology (Section \ref{rgbintegratedmassloss}) and hence justified the adopted gas-to-dust ratio of 400.}

\section{Pulsation effect on mass loss}
\subsection{Onset period of substantial mass loss for \kep\ stars}\label{onsetkeplerdata}
\mbox{\citetalias{yu2020a}} has confirmed the radial order assignment to the P--L sequences: sequence C stars pulsate in the fundamental mode ($n$=1), sequence C$^{\prime}$ and B in the first overtone (1O, $n$=2), and sequence A stars in the second overtone (2O, $n$=3), which is consistent with the findings by \citet{trabucchi2017a}.

Figure~\ref{fig:excesskepler} shows that substantial mass loss starts to occur at a pulsation period of $\sim$ 60 days, where the threshold of substantial mass loss is defined as \kwthree=0.3~mag, or equivalently \kwfour=0.55 mag, according to \citet{mcdonald2019a}. This onset period is associated with some of the LPVs pulsating in the first overtone ($n$=2, red circles), sequence C$^{\prime}$ and B, as shown in Figure~\ref{fig:excesskepler}a. Recently, \citet{mcdonald2019a} found that pulsators on sequences C and C$^{\prime}$ have substantial mass loss while stars on sequences B and A do not. This finding is confirmed by our results: Miras ($n$=1, pink asterisks) and  SRs pulsating in the first overtone ($n$=2, red circles) with longer periods exhibit substantial infrared excesses; First-overtone pulsators with shorter periods do not have substantial mass loss or, at least, their infrared colours are close to the thresholds. These stars are expected to be distributed along sequence B. Higher-overtone pulsators ($n \geqslant $3) have negligible mass loss. The critical period at $\sim$60~d is consistent with the results by \citet{mcdonald2016a}, who reported a similar period at $\sim$60~d using nearby RGB stars drawn from the Hipparcos catalogue.

Fig.~\ref{fig:excesskepler}(c) shows that a number of stars with periods $\rm P<20~days$ have infrared excesses, $K_s$-[22]>0.55~mag. This tendency is even more apparent in ASAS--SN and LMC LPVs, as shown later in Figs.~\ref{fig:periodamplitude}~and~\ref{fig:excessmagnitude}. To understand this, we first checked the photometric precision of the W4 magnitudes for these stars, and found that their W4 magnitudes have 48\% larger relative uncertainties than those of stars with $K_s$-[22] $\leqslant$ 0.55mag and periods $\rm P<20~days$. However, a typical (median) 2\% relative uncertainty in the W4 magnitude for the outliers cannot explain their significantly red colours. A possible reason could be related to target contamination. The \textit{WISE} photometry for all \kep\ and ASAS--SN stars in our sample have been examined in order to ensure that they are not contaminated or biased due to proximity to a known image artifact (by setting the quality flag \textit{ccf=\rm 0}). However, recently, \citet{dennihy2020a} used ground-based near-infrared and \textit{Spitzer} high-resolution imaging to show that \textit{WISE} targets that are not flagged for active deblending can suffer from unavoidable confusion for \textit{WISE}-selected infrared excesses. Hence, we think that these outliers are potentially a result of source confusion. 

Pulsation period and infrared colours, \kwthree\ and \kwfour, appear to be independent in higher-overtone pulsations ($n \geqslant $3), as revealed by the nearly horizontal trend in Fig.~\ref{fig:excesskepler}. \redbf{The remarkable match between the observed and modelled colour-period relations in the period range $P<$ $\sim$40 days revealed in Fig.~\ref{fig:excesskepler} confirms that the colour originates from the stellar photosphere rather than circumstellar envelope.} Fig.~\ref{fig:hrmodel} reveals that the scatter in the colours at a given period shown in Fig.~\ref{fig:excesskepler} is not caused by different masses. This is because at a given period the colours differ by a negligible amount, which is less than the formal errors of the colours, across different model masses. In fact, all seven models overlap in the \kwthree-period and \kwfour-period diagrams shown in Fig.~\ref{fig:excesskepler}.

\begin{figure*}
\begin{center}
\resizebox{0.7\textwidth}{!}{\includegraphics{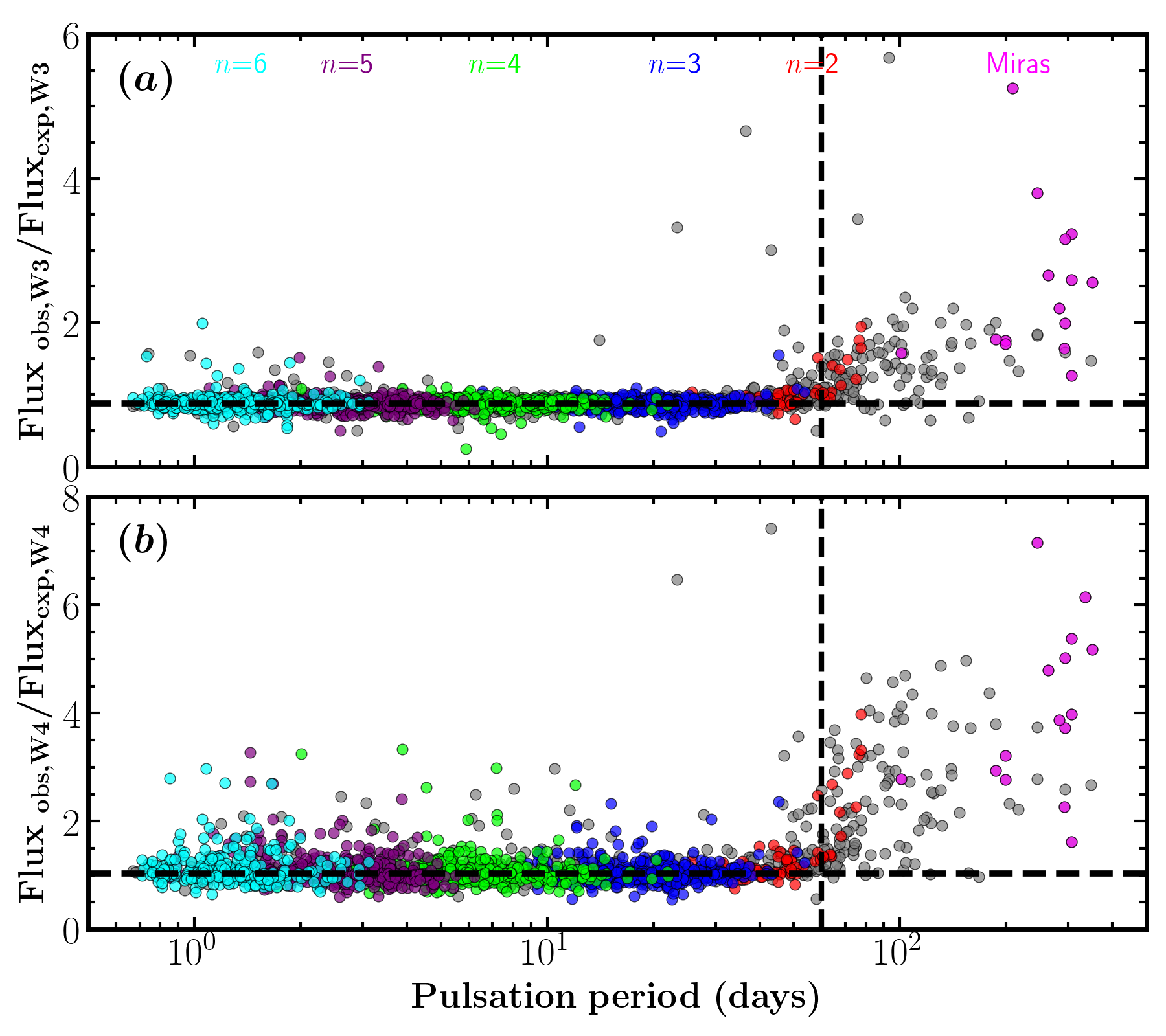}}\\
\caption{Ratios of the observed to expected flux densities, at the \textit{WISE} W3 (panel \textbf{a}) and W4 (panel \textbf{b}) bands as a function of pulsation period. The vertical dashed lines mark the period threshold, 60 days, of substantial mass loss, while the horizontal dashed lines define thresholds below which there is no dust production. Radial orders of the dominant modes are indicated at the top, except for the stars shown in the grey circles whose radial orders are unidentified due to short light curve coverage.} 
\label{fig:periodexcess}
\end{center}
\end{figure*}

According to the pulsation-enhanced dust-driven model, \redbf{we expect that pulsations trigger mass loss} because they lift material away from the star. Therefore, we might expect a sign that the onset of a dramatically increased mass loss coincides with a certain threshold in pulsation period, amplitude, and perhaps also in its energy. Fig.~\ref{fig:ampperiodpower} shows the mass-loss indicators as a function of pulsation amplitude, colour-coded by $\rm{log_{10}[(Amp/Period)^2]}$. Here the argument, $\rm{(Amp/Period)^2}$, is proportional to the power delivered by a sound wave. A rapid increase of mass loss is seen at the amplitudes of $\sim$50-100 mmag. This feature is similar to that seen in Fig.~\ref{fig:excesskepler}, due to the correlation between period and amplitude. A significant correlation between the mass-loss indicators and amplitude is visible for Miras (stars with Amp $>1$mag). There is a mass-loss gradient in the vertical direction in the amplitude range $\sim$50--100 mmag. This is because at a given amplitude a shorter period, thus less mass loss as per Fig.~\ref{fig:excesskepler}b, leads to a higher power indicated by the colour code. We also checked the correlations between the mass-loss indicators and other stellar parameters, i.e., luminosity and effective temperature, but no significant mass-loss increase occurring at a certain threshold was found when taking measurements uncertainties into account. Therefore, Figs~\ref{fig:excesskepler} and \ref{fig:ampperiodpower} appear to be in support of a direct link between pulsations and mass loss.

\begin{figure*}
\begin{center}
\resizebox{0.7\textwidth}{!}{\includegraphics{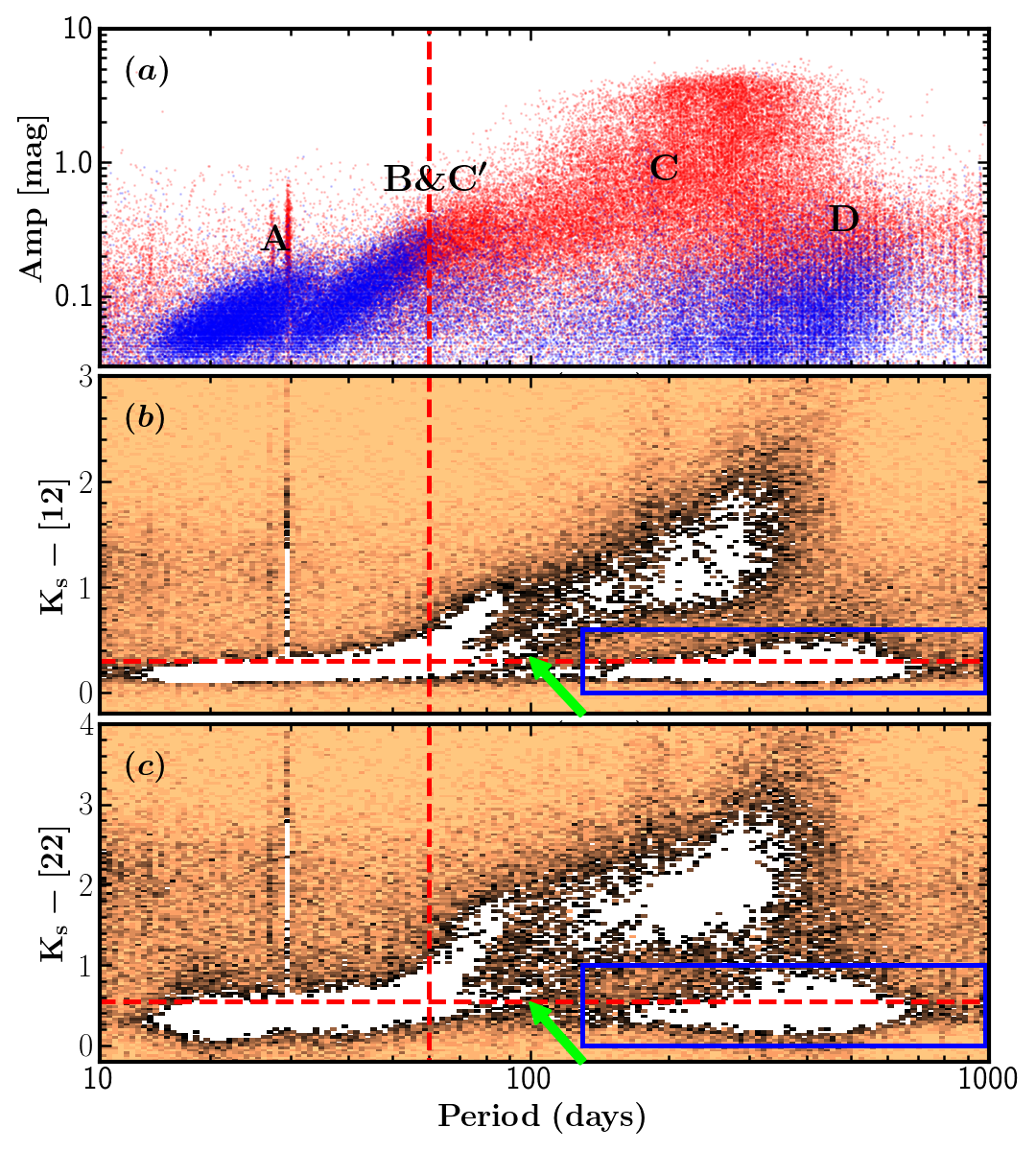}}\\
\caption{\textbf{(a)} Relation between pulsation periods and amplitudes of the dominant modes of ASAS--SN LPVS. The P--L sequences A, B, C$^{\prime}$, C, and D are indicated. LPVs shown in red are those with substantial mass loss (\kwthree >0.3 mag), while LPVs shown in blue are those without substantial mass loss (\kwthree\ $\leqslant$0.3 mag). This infrared-colour threshold defines the stars with dust production.\textbf{(b)} Infrared colour \kwthree\ against periods of ASAS--SN SRs (black) and Miras (red), and \kep\ (green) LPVs. The stars enclosed by the blue box are Long Secondary Period Variables. The green arrow indicates an onset of substantial mass loss at $\sim$100 days. \textbf{(c)} Same as panel \textbf{(b)} now for infrared colour \kwfour. The vertical dashed lines mark the period threshold, 60 days, of substantial mass loss, while the horizontal dashed lines define the threshold for stars producing dust (above) and not producing dust (below). Vertical stripes at 27 and 30 days in all panels are due to observational artifacts.} 
\label{fig:periodamplitude}
\end{center}
\end{figure*}

\subsection{Onset period of substantial mass loss by SED fitting}
An alternative way to estimate the mass-loss rate is using the ratio of the observed-to-expected flux densities at a given infrared band. We used the method detailed in Section~\ref{sed} to obtain expected flux densities based on the SED fitting. Fig.~\ref{fig:periodexcess} shows the ratio of the observed-to-expected flux densities in the \textit{WISE} W3 (upper panel) and W4 (lower panel) bands. The onset period of $\sim$60 days revealed by using infrared colours \kwthree\ and \kwfour\ in Section \ref{onsetkeplerdata} is reproduced in both bands. 

We note that there is a systematic offset of the flux ratios  in the \textit{WISE} W3 band in the sense that the observed flux densities are globally lower than the predicted flux densities by 11.8\%. In fact, a similar systematic offset was also found by \citet{mcdonald2009a}, who derived the flux ratios at 8$\mu$m. Although the origin of the offset has remained unclear, we think they are likely due to the zero point of the \textit{WISE} W3 photometry. Note that the flux extrapolation in the wavelength range 20--28~$\mu$m (see Section \ref{sed}) does not change the flux-ratio systematic offset in the W3 band, because the longest wavelength of the W3 band is shorter than 20$\mu$m. 

The systematic offset in the the \textit{WISE} W4 band is 3.5\% in the sense that the predicted flux densities are too low, but is much smaller than that seen in the W3 band. To test the influence of using different extrapolation methods on the flux ratios, we extrapolated a spectrum by fitting a Rayleigh-Jones tail rather than the \mbox{third-order} polynomial used in this work (see Section \ref{sed}), and found that it could not describe the spectrum. This is because the black-body spectrum deviates significantly from the synthetic spectrum for LPVs. Although we noticed that a second-order polynomial extrapolation gives a significant observed-to-expected flux densities offset (42.4\%) in the W4 band, the scatter of the flux ratio is hardly changed. Thus, to summarise, the relative rapid increase of the flux ratios at a period of $\sim$60 days is still apparent despite the systematic offsets.

\redbf{We note that synthetic spectra from hydrostatic models, which MARCS models are based on, may not properly describe the extended atmospheres of strongly pulsating LPVs. However, stars with the critical period of $\sim$60 days  should not have significantly extended atmospheres because they only lie just above the RGB tip \citep[e.g.,][]{kiss2003a, soszynski2004a, tabur2010a, wood2015a} and.}

\begin{figure}
\begin{center}
\resizebox{\columnwidth}{!}{\includegraphics{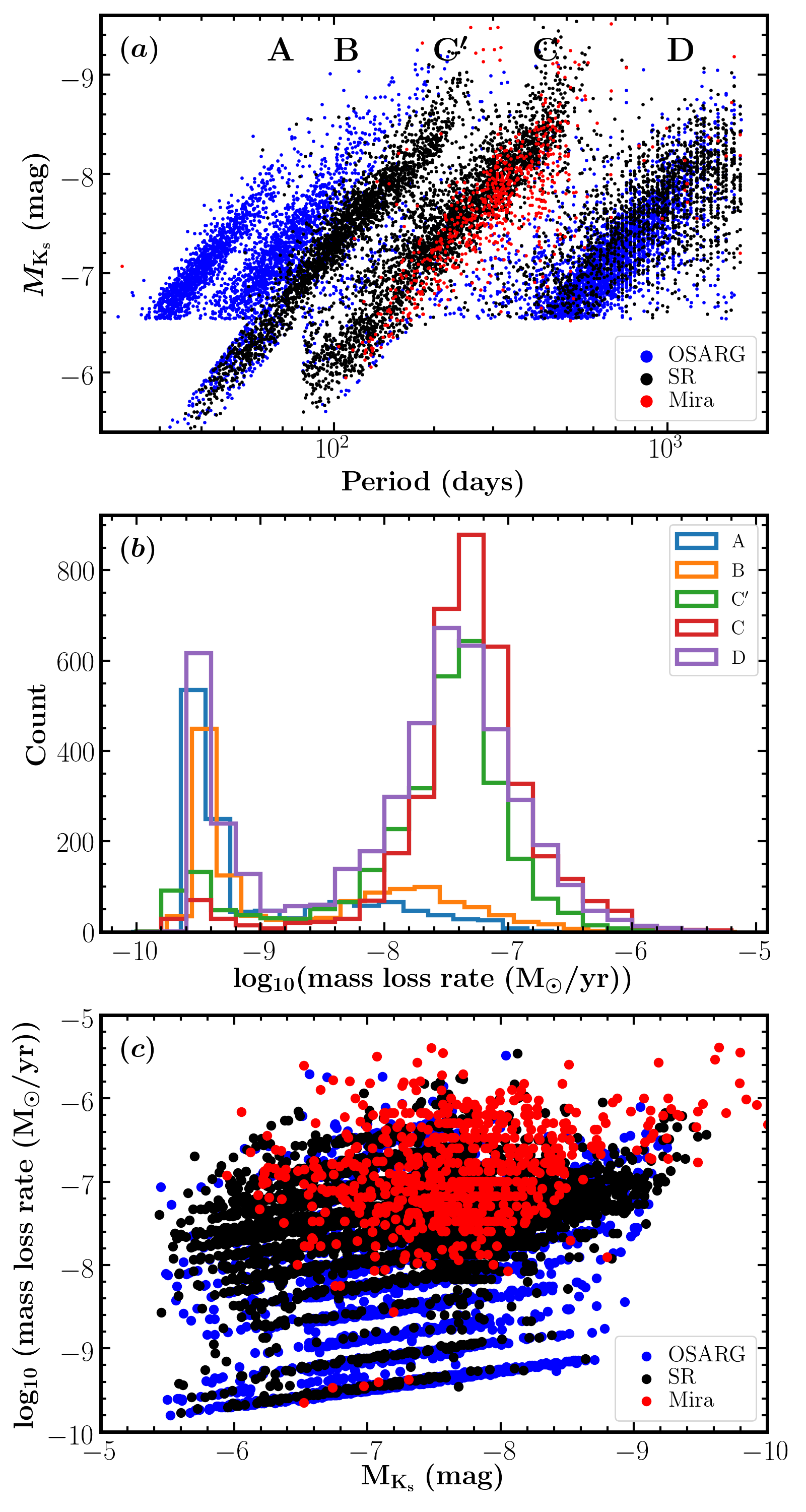}}\\
\caption{\textbf{(a)} Period-luminosity diagram of the LMC LPVs. Evolutionary phases of the SR (black), Mira (red), and OGLE Small Amplitude Red Giants (OSARG, blue) are taken from the OGLE-III catalogue of LPVs in the LMC \citet{soszynski2009a}. Absolute magnitudes, \mk, are derived using a distance modulus of 18.54 mag and an extinction correction of 0.0372 mag. Pulsation periods are adopted from \citet{riebel2012a}. Note that RGB stars are not included in this sample, leading to a sharp cutoff of the density of star number at \mk= --6.5 mag. \textbf{(b)} Distributions of the total (gas and dust) mass-loss rates of LPVs on sequences A, B, C$^{\prime}$, C, and SRs and OSARGs on sequence D. The total mass-loss rates are approximated by multiplying the dust mass-loss rates from \citet{riebel2012a} by a gas-to-dust mass ratio of 400.  \textbf{(c)} Mass-loss rate as a function of the absolute magnitude, \mk, for OSARGs (blue), SRs (black) and Miras (red). Stripes are a result of the relatively sparse GRAMS model grid used by \citet{riebel2012a}. The stars shown in the three panels are the same group of LMC LPVs.} 
\label{fig:massloss}
\end{center}
\end{figure}

\begin{figure}
\begin{center}
\resizebox{\columnwidth}{!}{\includegraphics{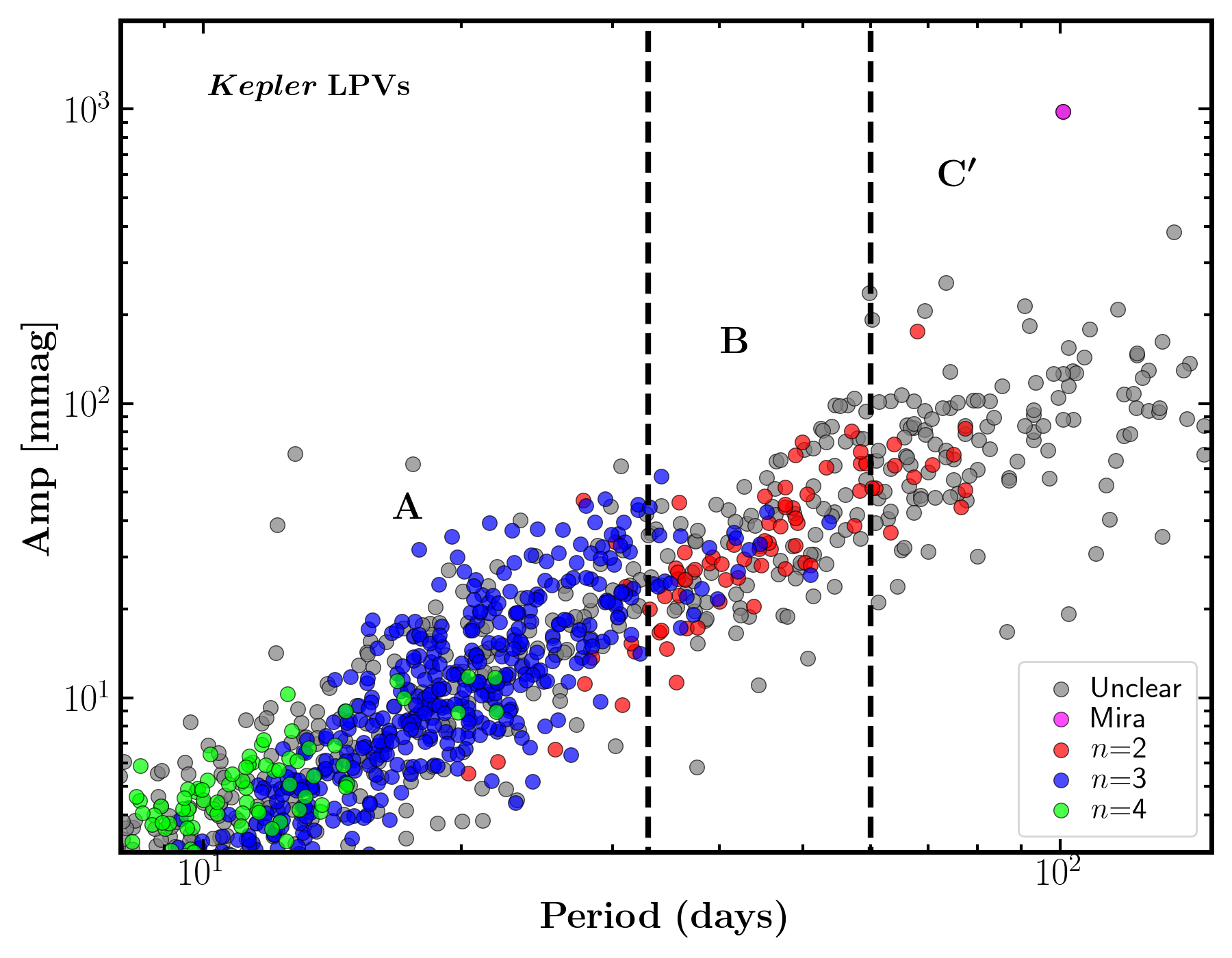}}\\
\resizebox{\columnwidth}{!}{\includegraphics{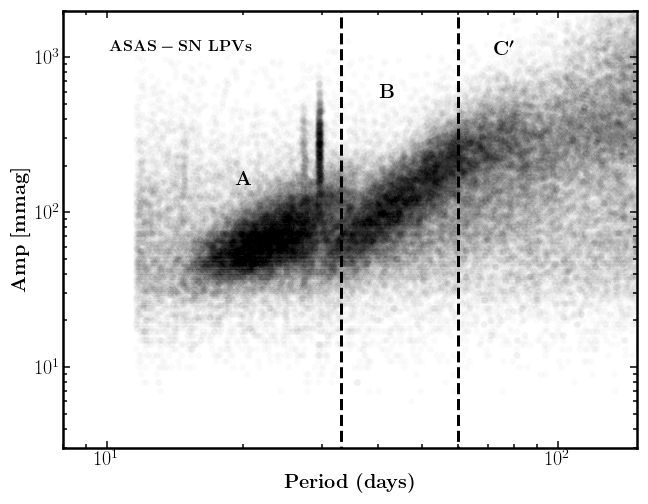}}\\
\resizebox{\columnwidth}{!}{\includegraphics{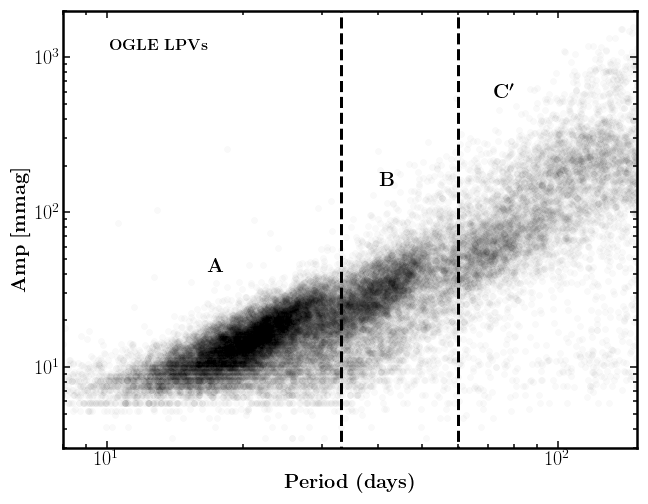}}\\
\caption{Relations of the pulsation periods and amplitudes of the dominant modes for \kep\ (\textbf{top}), ASAS--SN (\textbf{middle}), and OGLE (\textbf{bottom}) LPVs. In each panel, the left vertical dashed line marks an approximate period of the sequence transition from A to B, and the right vertical dashed line indicate the critical pulsation period at 60 days. The radial orders of the dominant modes are colour-coded and shown in the legend for \kep\ LPVs. Sequences A, B, and C$^{\prime}$ are shown for the respective regions in which stars of their sequences dominate. The pulsation periods and amplitudes of OGLE LPVs were taken from \citet{soszynski2009a}. Vertical stripes at 27 and 30 days in all panels are due to observational artifacts.}
\label{fig:PeriodAmp}
\end{center}
\end{figure}

By examining the relation between the ratios of the observed-to-expected flux densities in the \textit{WISE} W3/W4 bands and the pulsation amplitudes (not shown here), an almost identical trend as shown in Fig.~\ref{fig:ampperiodpower} was identified in Fig.~\ref{fig:periodexcess}, where the flux ratios rapidly increase with the pulsation amplitude at $\sim$50--100 mmag. This independent estimation of the mass-loss rate with the flux ratios confirms  the conclusion drawn from Fig.~\ref{fig:ampperiodpower} (see the previous Section).

\subsection{Onset period of substantial mass loss from ASAS--SN data}
From a combined sample of \kep\ and ASAS--SN LPVs, more features are revealed in Fig.~\ref{fig:periodamplitude}. Fig.~\ref{fig:periodamplitude}a shows amplitude versus period colour-coded by mass loss and looks similar to a the P--L diagram because amplitude scales with luminosity. We find that LPVs on sequence A generally have negligible infrared excess (blue points). Sequences B and C$^\prime$ merge together in the period-amplitude plane, where stars closer to sequence B on the shorter-period side appear to have no significant mass loss (blue points) while the stars closer to sequence C$^\prime$ with longer periods (red points) show substantial mass loss. This result is consistent with the findings by \citet{mcdonald2019a}.

A number of stars with periods in the range  $300~\rm{d} \lesssim P \lesssim 600~\rm{d}$ have low infrared excess (blue rectangles in Figs.~\ref{fig:periodamplitude}b and \ref{fig:periodamplitude}c and blue dots in Fig.~\ref{fig:periodamplitude}a). We note that some other stars with periods in the same range can have high mass-loss rates, and in general have larger pulsation amplitudes \citep[see Fig.~2 of][]{mcdonald2019a}.  These stars are likely to lie at the upper part of the region labelled by D in Fig.~\ref{fig:periodamplitude}a (red dots). All of the stars in these two groups have Long Secondary Periods and have been denoted as LSPs. This bimodal distribution of the mass loss in LSPs can be clearly seen in Fig.~\ref{fig:massloss}(b) (the purple distribution). \citet{trabucchi2017a} found that stars evolving from sequence B to C$^{\prime}$ develop into LSPs. The gap between sequence B to C$^{\prime}$ in the P--L plane seen in Fig.~\ref{fig:massloss}a can be filled by LSPs if the secondary peaks rather than the primary peaks of LSPs are used to construct a P--L diagram. While the cause of the LSPs is still unknown, pulsation is not a likely explanation \citep[e.g.,][]{nicholls2009a}. Thus, we confirm that some LSPs are more similar to LPVs on sequence C$^{\prime}$ and therefore have higher mass-loss rates, and other LSPs are more analogous to LPVs on sequence B and thus have lower mass-loss rates
(see Fig.~\ref{fig:massloss}b).

Fig.~\ref{fig:periodamplitude}(b) clearly shows a sudden increase in \kwthree\ occurring at a period of $\sim$ 60 days, consistent with the findings from \kep\ data (see Figs.~\ref{fig:excesskepler} and \ref{fig:lumicomp}). These stars are distributed along the merged sequences B and C$^{\prime}$ shown in Fig.~\ref{fig:periodamplitude}(a). \citetalias{yu2020a} has confirmed that the P--L sequences discovered in the Magellanic Clouds do not stand out so clearly in \kep\ LPVs, mainly because the typical uncertainty of \mk\ for \kep\ LPVs computed from Gaia DR2 parallaxes is approximately six times larger than that for LPVs in the LMC. We found that P--L sequences were also not presented in the P--L diagram of ASAS--SN LPVs. Therefore, we used LPVs in the LMC to understand if the onset is linked to stars on sequence B or C$^{\prime}$.

\begin{figure*}
\begin{center}
\resizebox{\columnwidth}{!}{\includegraphics{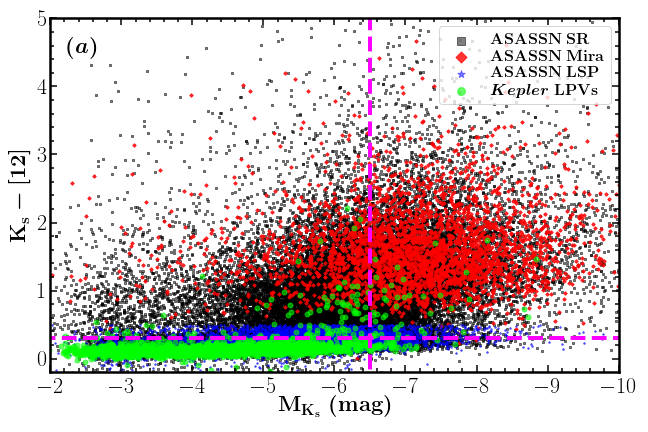}}
\resizebox{\columnwidth}{!}{\includegraphics{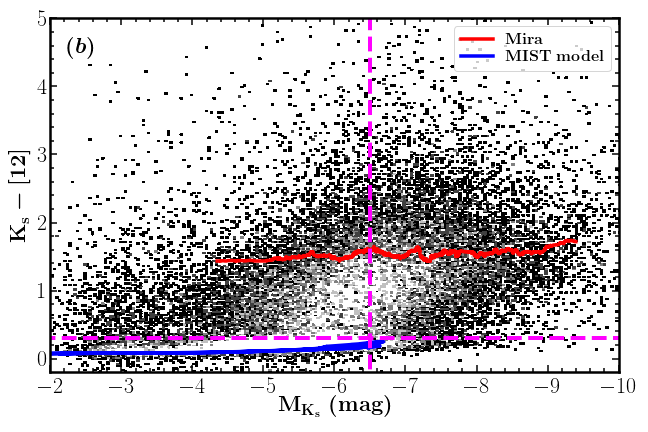}}\\
\resizebox{\columnwidth}{!}{\includegraphics{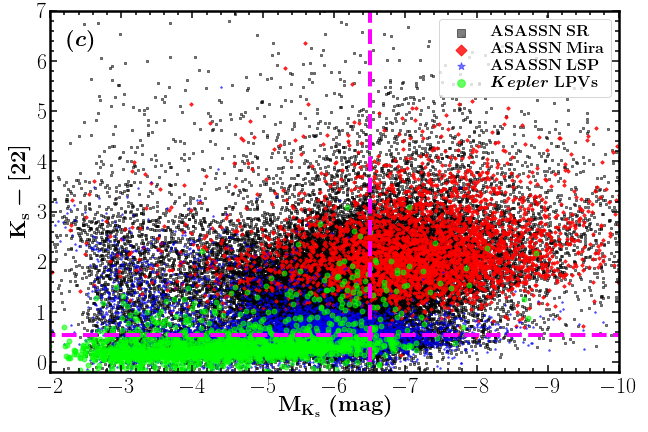}}
\resizebox{\columnwidth}{!}{\includegraphics{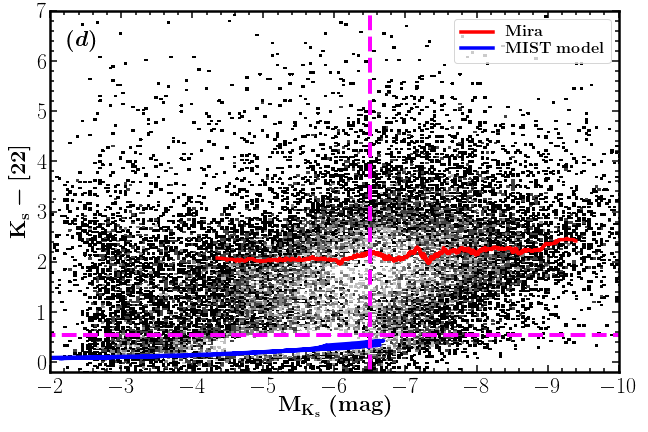}}\\
\caption{\textbf{(a)} Infrared colour \kwthree\ as a function of the absolute magnitude of the 2MASS Ks band, \mk, derived using Gaia DR2 parallaxes for the \kep\ and ASAS--SN LPVs. A parallax precision cut, $\sigma_{\pi}/\pi$<0.2, was applied. Overplotted are Miras (red), SRs (black), and Long Secondary Periods (blue) from the ASAS--SN survey, and the \kep\ LPVs (green). \textbf{(b)} Number density of all the stars shown in panel \textbf{(a)}. The red-solid curve indicates the moving median colour \kwthree\ as a function of the absolute magnitude \mk\ for the Miras, while the blue-solid line shows the relation between \kwthree\ and \mk\ for the RGB, both of which were calculated using dust-free \redbf{MIST models}.  \textbf{(c)} and \textbf{(d)} similar as panels \textbf{(a)} and \textbf{(b)} now using infrared colour \kwfour. The vertical dashed lines mark an approximate absolute magnitude at the tip of the RGB for the LMC LPVs, while the horizontal dashed lines define the threshold for stars producing dust (above) and not producing dust (below). Note that the significant mass loss sets in for fainter ASAS--SN LPVs, $\rm -5.0~mag<\mk<-6.4~mag$, than for the LMC LPVs, \mk$\simeq$--6.5~mag. This implies that those ASAS--SN LPVs have higher masses (see the text in Section~\ref{radiationmassloss} for more details).}
\label{fig:excessmagnitude}
\end{center}
\end{figure*}

Fig.~\ref{fig:massloss}(b) shows that LPVs on sequence A and B do not have appreciable mass-loss rates, which are estimated to be $5\times10^{-10}$\msun/yr, while LPVs on sequence C$^{\prime}$ exhibit high mass-loss rates, which peak at $6\times10^{-8}$\msun/yr. The dust mass-loss rates were calculated for the LPVs in the LMC, based on radiative transfer models of evolved stars and circumstellar dust shells and by assuming a constant expansion velocity. We refer the reader to \citet{riebel2012a} and references therein for more details. The significant difference of the mass loss rates, by two orders of magnitude, could imply that the onset at 60 days is linked to the base of sequence C$^{\prime}$. 

% This implication should not be impacted by RGB stars, which are not included in the sample but are mainly located along sequence A and B and below the artificial cutoff \mk=-6.5 mag (see Fig.~\ref{fig:massloss}a). These excluded RGB stars should have lower mass-loss rates than the AGB stars higher up along the same sequences. This is because the RGB stars have lower luminosites, which leads to lower mass-loss rates given the strong correlation between \mk\ and mass-loss rates for less luminous stars (see Fig.~\ref{fig:massloss}c).

Fig.~\ref{fig:periodamplitude}(b) also suggests another threshold of substantial mass loss at $\sim$100 days, indicated by the green arrow. To understand whether this critical period is linked to the base of sequence C, we binned the periods in steps of 5 days and calculated the mean periods and mean mass-loss rates for the LPVs on sequence C in the LMC shown in Fig.~\ref{fig:massloss}(b). We found a clear increase of the mean mass-loss rates from 1.6$\times10^{-8}$ to 8.4$\times10^{-8}$\msun\//yr with periods from 80 to 120 days. More importantly, we found a similar increase of the mean mass-loss rates from 1.6$\times10^{-8}$ to 7.0$\times10^{-8}$\msun\//yr with periods from 45 to 95 days for the LPVs on sequence C$^{\prime}$ in the LMC. This similarity is in favor of the critical period of $\sim$100 days being linked to the base of sequence C and the critical period of $\sim$60 days being linked to the base of sequence C$^{\prime}$. Fig.~\ref{fig:periodamplitude}(c) shows clearly the first onset at $\sim$ 60 days, but less clearly the second onset at $\sim$ 100 days, which could be explained by more precise photometry in the \textit{WISE} W3 band than the W4 band.

Figs.~\ref{fig:periodamplitude}(b) and \ref{fig:periodamplitude}(c) show that mass-loss rates plateau at \kwthree~$\approx$ 1 mag or \kwfour~$\approx$ 2 mag in the pulsation period range $\mbox{100 d \la \textit{P} \la 300 d}$. For these stars, the typical mass-loss rate was estimated to be $\sim8.6\times10^{-8}$\msun/yr, corresponding to the right peak of the bimodal distribution in Fig.~\ref{fig:massloss}(b). Mass-loss rates resume to increase in Miras (see red dots in Fig.~\ref{fig:periodamplitude}b and ~\ref{fig:periodamplitude}c).

It would be reasonable to expect that the pulsation amplitude increases dramatically at the critical pulsation period of $\sim$60 days, given that mass loss is enhanced by pulsations. However, this expectation is not clearly visible with \kep, ASAS--SN, and OGLE LPVs shown in Fig.~\ref{fig:PeriodAmp}. Considering the presence of a rapid increase of mass loss at certain thresholds of period and the absence of the rapid increase of amplitude at the same threshold of period, it would suggest that the link between pulsation and mass-loss is not simple. A systematic decrease in amplitude is seen around the sequence transition from A to B from each of the three groups of LPVs.

\begin{figure*}
\begin{center}
\resizebox{1.0\textwidth}{!}{\includegraphics{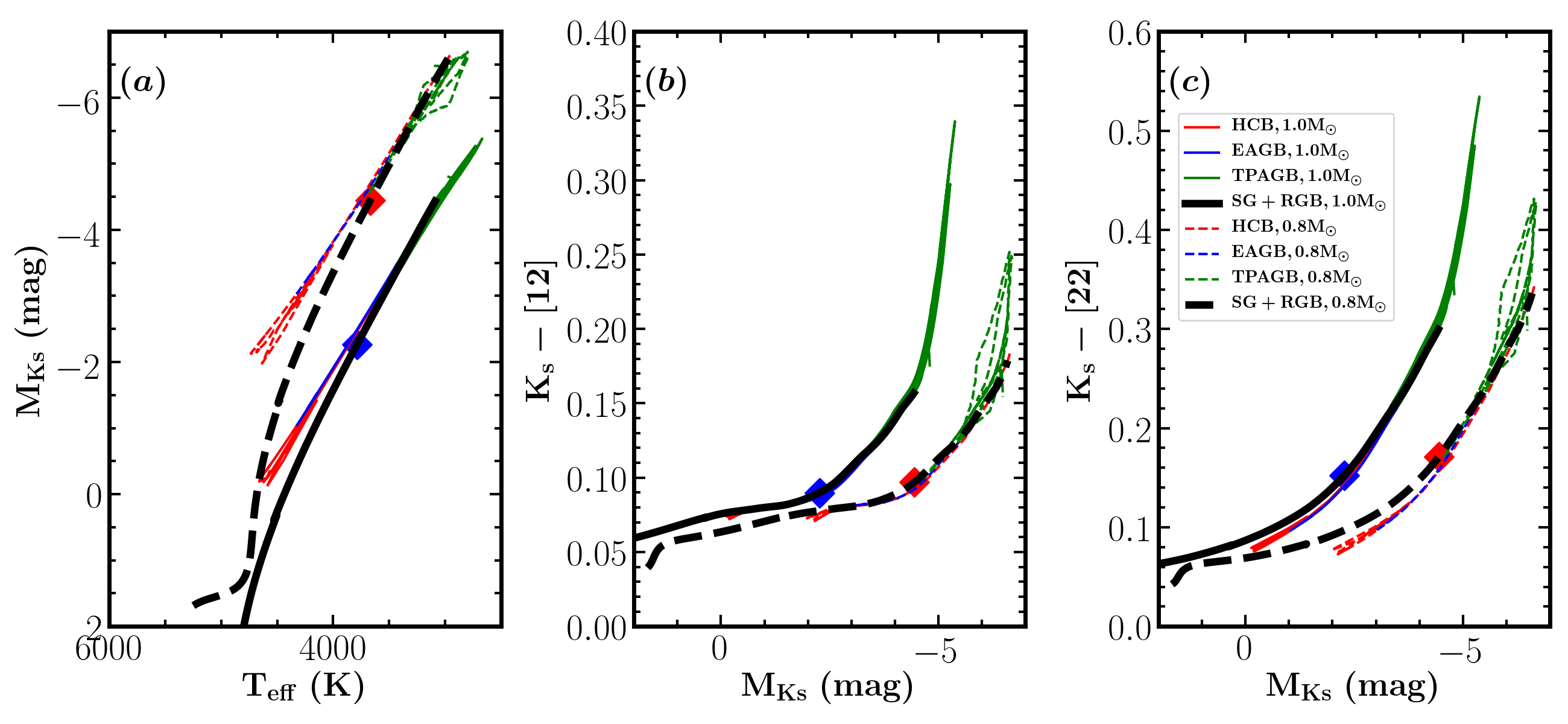}}\\
\caption{(\textbf{a}) Evolutionary tracks of different masses (0.8 and 1.0 \msun) at solar metallicity, adopted from \redbf{MIST database}. Evolutionary phases range from the subgiant to the thermally pulsing AGB,  indicated with different colours as given in the legend of panel \textbf{c}. The absolute magnitude, \mk, is derived from MIST model luminosities and bolometric correction. (\textbf{b}) Dust-free model infrared colours \kwthree\ and  (\textbf{c}) \kwfour\ as a function of the absolute magnitude, \mk. The two diamonds defined in Fig.~\ref{fig:hrmodel}a highlight the same luminosity but the significantly different \mk, due to the difference in \teff.}  
\label{fig:abMag}
\end{center}
\end{figure*}

\section{Effect of Radiation on mass loss}
\label{radiationmassloss}
\redbf{We now focus the investigation on the effect of radiation on mass loss for Miras and SRs.  Fig.~\ref{fig:excessmagnitude} shows that for Miras (red dots in panels \textbf{a} and \textbf{c}), mass-loss rates do not depend on radiation, since \kwthree\ and \kwfour\ are approximately constant with \mk. This can be explained by the fact that the radiation pressure (luminosity) mainly determines the wind velocity, while the density in the dust formation zone largely determines the mass-loss rate (note that the mass-loss rate is proportional to both the velocity and the density). Typical values of the wind velocity span about one order of magnitude on the AGB, while the observed mass-loss rates for Mira span several orders of magnitude (for both observations and dynamical models, see Fig. 7 of \citealt{bladh2019b} and references therein).}

To investigate further how mass-loss rates vary with \mk\ in SRs, we show the number density of stars in Figs.~\ref{fig:excessmagnitude}b and \ref{fig:excessmagnitude}d. Apparently, substantial mass-loss in SRs sets in below the RGB tip of the LMC LPVs (marked by the vertical dashed lines), increases rapidly with decreasing \mk\ (namely with increasing luminosity), and eventually plateaus at a level similar to the typical mass-loss rate in Miras (almost horizontal lines in red). Moreover, infrared colours \kwthree\ and \kwfour\ increase with decreasing \mk\ in lower-luminosity \kep\ red giants (green dots). 

It is worth noting that the variations of mass-loss rates revealed by the infrared colours for the \kep\ and ASAS--SN LPVs (Figs.~\ref{fig:excesskepler} and \ref{fig:periodamplitude}) are consistent with the variations of mass-loss rates modelled by solving radiative transfer equations for the LPVs in the LMC (Fig.~\ref{fig:massloss}c).  Fig.~\ref{fig:massloss}c shows that Miras (red dots) exhibit essentially constant mass-loss rates, while SRs (black dots) exhibit increasing mass-loss rates at fainter magnitudes (left side of Fig.~\ref{fig:massloss}c), and eventually flattens off toward brighter magnitudes (right side of Fig.~\ref{fig:massloss}c) at the same mass-loss-rate level as Miras. Most lower-luminosity red giants (included in OSARGs in Fig.~\ref{fig:massloss}c) have low mass-loss rates below $\rm 10^{-8}M_{\odot}/yr$ that are correlated with \mk. 

\redbf{The infrared colour versus \mk\ relation shown in Fig.~\ref{fig:massloss} is also in line with observations and dynamical models by \citet{bladh2019b}. They found that the mass-loss rate increases with luminosity and with period, and progressively flattens off (see their Figs. 8 and 13).} The variations of mass-loss rates with \mk\ are actually consistent with the pulsation-enhanced dust-driven model, because the mass-loss rates are presumably determined by pulsations through levitating the upper atmosphere into the circumstellar envelope, and the terminal outflow velocities are presumably determined by radiation pressure \redbf{(for models, see \citealt{mattsson2010a,eriksson2014a,bladh2019b}; for observations, see \citealt{mcdonald2019b}; and for a recent review see \citealt{hofner2018a}.)}

The RGB tip of the ASAS--SN LPVs appears fainter (higher \mk) than that of the LMC LPVs, and this is likely the result of differing masses. We note that \mk\ = --6.5 mag has been quoted for the LMC LPVs near the RGB tip \citep[e.g., see][]{tabur2010a, yu2020a}. This value is also consistent with the prediction from the 0.8\msun\ model (see in Figs.~\ref{fig:excessmagnitude}b and \ref{fig:excessmagnitude}d and Fig.~\ref{fig:abMag}). Fig.~\ref{fig:abMag} shows that the absolute magnitude of the RGB tip depends on mass, where stars that evolve to the RGB tip appear fainter if they have higher masses. Keeping in mind that the threshold period of substantial mass loss occurs at $\sim$60 days and is just above the periods of the RGB tip,  the rapid increase in the mass-loss rates of ASAS--SN LPVs on the fainter side $\rm -6.4~mag<\mk<-5.0~mag$ suggests the mean mass of ASAS--SN LPVs near the RGB is higher than the mean mass of their counterparts in the LMC.

% \begin{figure}
% \begin{center}
% \resizebox{\columnwidth}{!}{\includegraphics{Figures/Age.png}}\\
% \caption{Absolute magnitude, $M_{\rm K_{s}}$, as a function of stellar age for 0.8 (blue dashed line) and 1.0 \msun\ (red solid line) models along the RGB. The age of the RGB base is taken as reference. For each model, the shaded region marks the evolutionary phase from the luminosity bump to the tip of the RGB, with the lifetime indicated.} 
% \label{fig:age}
% \end{center}
% \end{figure}

\begin{figure*}
\begin{center}
\resizebox{0.9\textwidth}{!}{\includegraphics{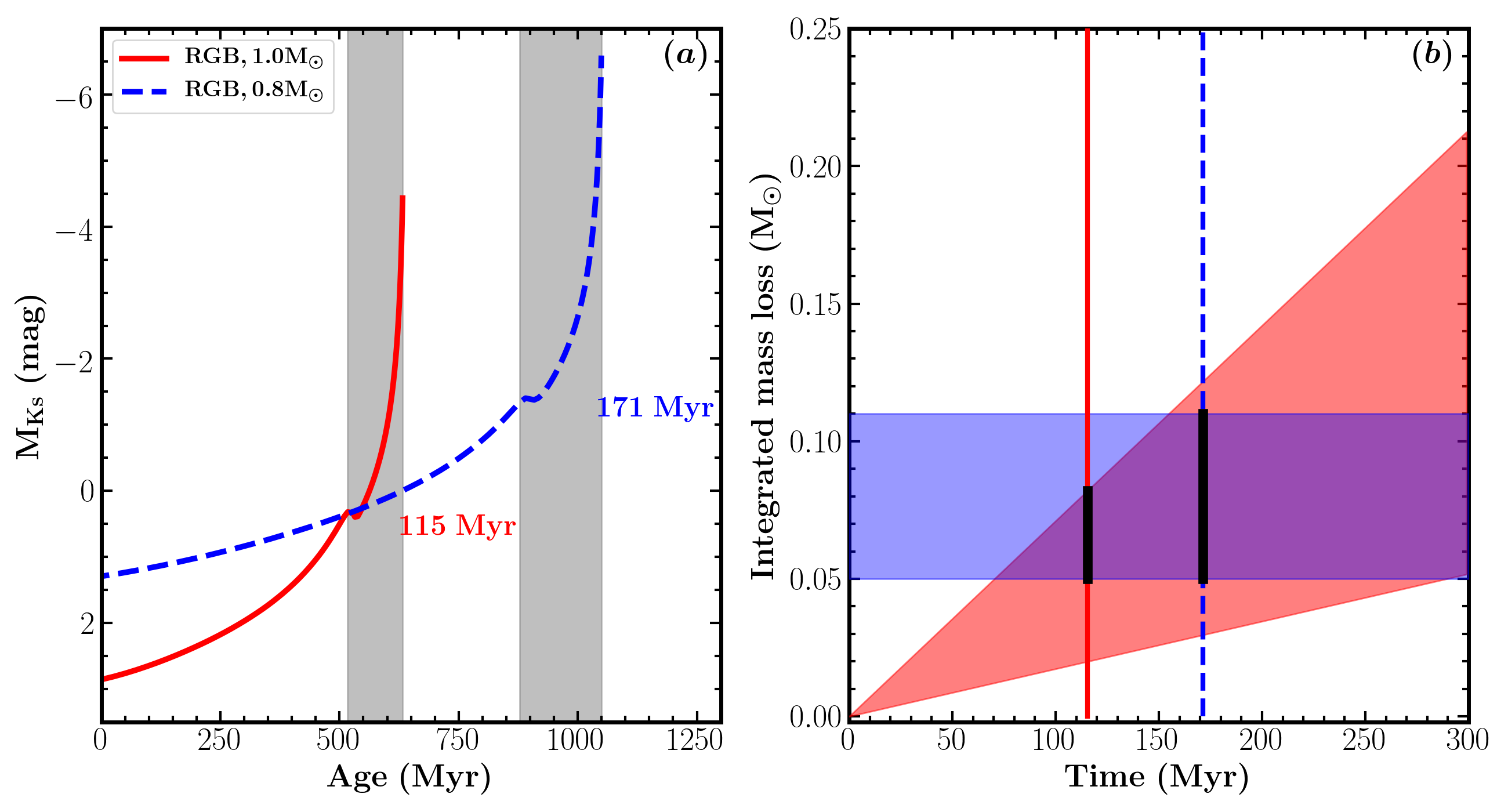}}\\
\caption{\textbf{(a)} Absolute magnitude, $M_{\rm K_{s}}$, as a function of stellar age for 0.8 (blue dashed line) and 1.0 \msun\ (red solid line) models along the RGB. The age of the RGB base is taken as reference. For each model, the shaded region marks the evolutionary phase from the luminosity bump to the RGB tip, with the lifetime indicated. \textbf{(b)} Integrated mass loss as a function of time, over which the integrated mass loss is evaluated, for the 0.8 (vertical blue dashed line) and 1.0 \msun\ (vertical red solid line) models along the RGB. Note that the age of the luminosity bump is taken as reference for each model. The red shaded region (triangle) shows the integrated mass loss estimated from this work by estimating a mass-loss rate of $4.42 \times 10^{-10}\pm2.69 \times 10^{-10}$\msun/yr, while the blue shaded region (rectangle) represents the integrated mass loss measured from asteroseismology, $\Delta M=0.08 \pm 0.03$\msun \citep{kallinger2018a}. The two thick vertical bars indicate the consistency of the integrated mass loss measured from the two methods.} 
\label{fig:age}
\end{center}
\end{figure*}

\section{Integrated Mass Loss on the Red Giant Branch}\label{rgbintegratedmassloss}
Various studies have shown that the pulsation periods of the primary modes for stars near the RGB tip span a range 20-50 days \citep[e.g.,][]{kiss2003a, soszynski2004a, tabur2010a, wood2015a}. The upper limit of this range is near the threshold period of $\sim$ 60 days above which we see substantial mass loss. Thus, the mass-loss rate of RGB stars near the RGB tip should be relatively low. In fact, these stars are expected to be mainly located along sequences B and A \citep{soszynski2004a}. The typical (mean) mass-loss rate for the LMC stars on sequences B and A is estimated to be $4.42 \times\ 10^{-10}\pm2.69 \times\ 10^{-10}$ \msun/yr (Fig.~\ref{fig:massloss}b). If we approximate the lifetime of the RGB phase from the luminosity bump to the tip for a 1.0 \msun\ star to be 115 Myr, the expected integrated mass loss on the RGB is typically 0.05$\pm$0.03 \msun (see Fig.~\ref{fig:age}). Similarly, if we approximate the lifetime for a 0.8 \msun\ star to be 171 Myr, the integrated mass loss is 0.08$\pm$0.05\msun (see Fig.~\ref{fig:age}).

Interestingly, these estimates of the integrated mass loss are consistent with the findings from asteroseismic analysis (see Fig.~\ref{fig:age}b). \citet{miglio2012a} investigated solar-like oscillations of 40 stars on the RGB and 19 in the red clump of the old \mbox{metal-rich} cluster NGC 6791, and derived the difference between the average mass of RGB and red-clump stars from the seismic scaling relations. They found a difference of $\Delta M=0.09 \pm 0.03$ (random)$\pm$ 0.04 (systematic) \msun. \citet{kallinger2018a} carried out a detailed analysis of solar-like oscillations in eclipsing binary systems and calibrated the seismic scaling relations. They recalculated the difference between the average mass of RGB and red-clump stars using their non-linear scaling relations and obtained similar result, which is $\Delta M=0.08 \pm 0.03$ (random) \msun. Recently, \citet{miglio2020a} used a grid-based modelling method to estimate the integrated mass loss and confirmed the previous asteroseismic estimates by reporting $\Delta M=0.09 \pm$ 0.05 \msun\ for stars in the cluster NGC 6791 and $\Delta M=0.10 \pm$ 0.02 \msun\ for a high-$\alpha$ population of oscillating red giants. 

\redbf{We note that the typical (mean) mass-loss rate estimated here for the LMC stars on sequences A and B ($4.42 \times\ 10^{-10}\pm2.69 \times\ 10^{-10}$ \msun/yr) is based on our assumptions of the wind velocity (10 km/s) and the gas-to-dust mass ratio (400, see Section~\ref{gridmodels}). The good consistency between the integrated mass loss estimated from this work and from asteroseismology justifies these assumptions.}

The integrated mass loss, if taken to be  $\Delta M=0.08$~\msun, is only 6.3\% of the typical (mean) mass of red-clump stars $M=1.26$~\msun, and within 1$\sigma$ formal uncertainties on a single star basis $\sigma_{\rm M}=0.13$~\msun, according to the seismic masses of 16,000 \kep\ red giants by \citet{yu2018a}. Helium-core burning stars were distinguished from RGB stars therein via asteroseismology. Therefore, based on current precision of mass estimates from asteroseismology, we conclude that mass loss is currently not a limitation of stellar age estimates for galactic archaeology studies.

\section{Conclusions}
We investigated mass loss in a sample of 3233 luminous \kep\  red giants and 135,928 ASAS--SN SRs and Miras, using infrared colours (\kwthree\ and \kwfour) and ratios of the observed-to-expected flux densities at the \textit{WISE} W3 and W4 bands to estimate mass-loss rates. We explored the interplay between mass loss, pulsations and radiation, and compared our findings with the result based on LPVs in the LMC, to which precise distances have previously been determined. The typical integrated mass loss on the RGB was estimated and compared to asteroseismic analysis of \kep\ cluster stars exhibiting solar-like oscillations. Our findings can be summarised as follows:
\begin{itemize}
    \item The pulsation-enhanced dust-driven mass loss sets in at a pulsation period of $\sim$60 days. This critical period corresponds to the pulsations in the first overtone, namely $n=$2. The onset period is also revealed by the infrared colours \kwthree\ and \kwfour\ and flux ratios based on SED fitting. This result is consistent with previous studies. This rapid increase of mass loss occurs at pulsation amplitudes of $\sim$50-100 mmag. We find a second critical period at $\sim$ 100 days. The onset period around 60 days, which is just above the RGB tip, approximately corresponds to the base of P--L sequence C$^{\prime}$, while the onset period around 100 days corresponds to the base of P--L sequence C. Stars on both sequences exhibit high mass-loss rates with a mean value of $8.6\times10^{-8}$\msun/yr.
    
    \item Our analysis clearly reveals that the mass-loss rate in Miras does not depend on radiation. \redbf{This result is consistent with the mass loss in LPVs being pulsation-enhanced dust-driven outflows}: it seems that mass loss is set by the levitation of upper atmosphere into the circumstellar envelope by pulsations, and terminal outflow velocity is set by radiation. The mass-loss rate in SRs increases rapidly with luminosity, and eventually plateaus to a level similar to that in Miras. We find that the masses of ASAS-SN stars in the vicinity of the RGB tip may be higher than the masses of their counterparts in the LMC. 
    
    \item \redbf{The typical mass-loss rate in luminous RGB stars is estimated to be $4.42 \times 10^{-10}\pm2.69 \times 10^{-10}$ \msun/yr by assuming a wind velocity of 10 km/s and a gas-to-dust mass ratio of 400}. By adopting a representative 171 Myr lifetime from the luminosity bump to the RGB tip for a 0.8\msun\ star, the typical integrated mass loss on the RGB is estimated to be 0.08$\pm$0.05~\msun. This is consistent with the mass loss measured from asteroseismology of red giants in open clusters (0.09$\pm$0.05\msun\ integrated mass loss estimated by \citealt{miglio2012a, miglio2020a}, and 0.08$\pm$0.03\msun\ estimated by \citealt{kallinger2018a}).
\end{itemize} 

Recently, \citet{cunha2020a} suggested that the transition of pulsations from being predominantly stochastically driven to being predominantly coherently driven occurs around pulsation periods of 60 days. This critical period is equal to the observed onset period of substantial mass loss, and hence provides a new clue for understanding how the pulsation-enhanced dust-driven mechanism sets in and works in detail. The higher-precision Gaia parallaxes and the longer coverage of ASAS-SN light curves to be released in the future will refine the measurements of period and absolute magnitude, which will enable us to better understand the interplay between mass loss and pulsations \& radiation. \textit{TESS} is observing far more stars in open clusters than \textit{Kepler}, and \redbf{asteroseismology of these stars will offer a new opportunity for measuring integrated mass loss in RGB stars by comparing the average masses of red-clump and RGB stars.}

\section*{Acknowledgements}
We thank the anonymous referee for very helpful comments. We also thank Andrea Miglio for his feedback after our manuscript was put on the KASOC. J.Y. and L.G. were partially  supported by PLATO grants from the Deutsches Zentrum f\"{u}r Luft- und Raumfahrt (DLR 50OO1501) and from the Max Planck Society. The computational resources were provided by the DLR-funded German Data Center for SDO. J.Y. and S.L.B. acknowledges the Joint Research Fund in Astronomy (U2031203) under a cooperative agreement between the National Natural Science Foundation of China (NSFC) and Chinese Academy of Sciences (CAS). T.R.B. acknowledges support from the Australian Research Council, and from the Danish National Research Foundation (Grant DNRF106) through its funding for the Stellar Astrophysics Center (SAC). D.H. acknowledges support from the Alfred P. Sloan Foundation and the National Aeronautics and Space Administration (80NSSC19K0597). 

\section*{Data Availability}
The data underlying this article will be shared on reasonable request to the corresponding author.

%%%%%%%%%%%%%%%%%%%%%%%%%%%%%%%%%%%%%%%%%%%%%%%%%%

%%%%%%%%%%%%%%%%%%%% REFERENCES %%%%%%%%%%%%%%%%%%

% The best way to enter references is to use BibTeX:

\bibliographystyle{mnras}
\bibliography{reference.bib} % if your bibtex file is called example.bib

% Don't change these lines
\bsp % typesetting comment
\label{lastpage}
\end{document}